\documentclass[aps,prd,preprint,superscriptaddress,showkeys,amsmath,amssymb]{revtex4-2}

\usepackage{graphicx}
\usepackage{braket}
\usepackage{bm}

\newcommand{\nd}{\mathrm{d}}

\newcommand{\lbra}[1]{\langle #1 |}
\newcommand{\rbra}[1]{| #1 \rangle}

\newcommand{\prop}{\text{prop}}
\newcommand{\isp}{\text{isp}}

\begin{document}


\title{The recursive structure of Baikov representations: the top-down reduction with intersection theory}


\affiliation{Zhejiang Institute of Modern Physics, School of Physics, Zhejiang University, Hangzhou 310027, China}
\affiliation{CAS Key Laboratory of Theoretical Physics, Institute of Theoretical Physics, Chinese Academy of Sciences, Beijing 100190, China}

\author{Xuhang Jiang}
\email{xhjiang@itp.ac.cn}
\affiliation{CAS Key Laboratory of Theoretical Physics, Institute of Theoretical Physics, Chinese Academy of Sciences, Beijing 100190, China}
\author{Ming Lian}
\email{mlian@zju.edu.cn}
\author{Li Lin Yang}
\email{yanglilin@zju.edu.cn}
\affiliation{Zhejiang Institute of Modern Physics, School of Physics, Zhejiang University, Hangzhou 310027, China}



\date{\today}

\begin{abstract}
    Following our previous study of the recursive structure of Baikov representations, we discuss its application in the integration-by-parts reduction of Feynman integrals. We combine the top-down reduction approach with the recursive structure, which can greatly simplify the calculation for each sector in many cases. We introduce a new concept called the top-sector ISP reduction, which generalizes the maximal-cut reduction by retaining the sub-sector information. After subtracting the top-sector components, we provide a general method to transform the remaining integrand explicitly to sub-sectors, such that the reduction procedure can be carried out recursively. In this work, we use the intersection theory to demonstrate our method, although it can be applied to any implementation of the integration-by-parts reduction.
\end{abstract}

\keywords{Feynman integrals, Baikov representations, intersection theory}

\maketitle

\section{Introduction}
	
	Feynman integrals (FIs) are building blocks of perturbative scattering amplitudes in quantum field theories. In the calculation of a particular scattering amplitude, one often encounters a huge number of FIs. To compute them, one reduce them to a basis called Master Integrals (MIs). The number of MIs is much smaller. They can then be calculated using various methods, in particular, the method of differential equations \cite{Kotikov:1990kg, Henn:2013pwa, Henn:2014qga}.
	
	In practice, the reduction of FIs usually proceeds by solving Integration-By-Parts (IBP) relations among different integrals. These relations form a linear system that can be solved by the Laporta algorithm \cite{Laporta:2000dc, Laporta:2000dsw}. This IBP reduction procedure has been implemented in several public packages such as \texttt{Reduze} \cite{vonManteuffel:2012np}, \texttt{LiteRed} \cite{Lee:2013mka}, \texttt{FIRE} \cite{Smirnov:2019qkx} and \texttt{Kira} \cite{Klappert:2020nbg}. Recently, a novel method, the intersection theory \cite{Mizera:2017rqa, Mastrolia:2018uzb, Mizera:2019ose, Frellesvig:2019uqt, Frellesvig:2020qot, Weinzierl:2020xyy, Caron-Huot:2021xqj, Caron-Huot:2021iev, Fontana:2023amt, Chestnov:2022xsy}, is proposed to perform the integral reduction using the language of twisted cohomology groups. This regards IBP equivalence classes of Feynman integrals as elements in a cohomology group, and uses a concept called intersection numbers to compute the reduction coefficients.
	
	The IBP systems can become very large in cutting-edge applications. Generating and solving the relations is often a major bottleneck in multi-loop calculations. It is therefore desirable to reduce the size of the IBP system as much as possible. For example, packages like \texttt{NeatIBP} \cite{Bohm:2018bdy, Wu:2023upw} and \texttt{Blade} \cite{Liu:2018dmc, Guan:2019bcx} have been developed to achieve this goal by pre-selecting a smaller set of IBP relations before performing the full reduction. Another way to reduce the size of the IBP system is to split it into smaller subsystems, which can be solved separately and glued together for the final results. To this end, generalized unitarity cuts of Feynman integrals provide a powerful tool. Under a certain cut, a lot of integrals vanish and drop out of the linear relations, effectively making the system smaller. In the literature, there are two kinds of approaches to take advantage of cuts: the ``bottom-up'' approach and the ``top-down'' approach. In the bottom-up approach, one chooses a set of ``spanning-cuts'' which is a minimal set of cuts necessary to recover the full information. The reduction is performed under each cut in the set, and the full results are then assembled from these partial ones.
	
	The top-down approach \cite{Kardos:2018uzy}, on the other hand, starts from the top-sector containing the maximal number of propagators in a given integral family. By imposing the maximal cut, i.e., localizing all propagator denominators to the mass shells, it is easy to compute the reduction coefficients in the top-sector. One then subtract the top-sector components from the integrals to be reduced, and move to sub-sectors with fewer propagators. Recursively applying the above procedure down to the lowest sectors, one achieves the full reduction of the integrals.
	
	In a recent work \cite{Jiang:2023qnl}, we have explored the recursive structure of Feynman integrals, which is particularly apparent in the Baikov representations \cite{Baikov:1996iu, Lee:2009dh}. The Baikov representations of Feynman integrals amount to a change of integration variables from loop momenta to propagator denominators. As a result, it is rather straightforward to study cuts of integrals in these representations. Imposing a cut is simply taking the residue at the origin for a variable \cite{Frellesvig:2017aai, Bosma:2017ens, Harley:2017qut}. IBP relations can also be studied in the Baikov representations \cite{Larsen:2015ped, Bohm:2018bdy, Kardos:2018uzy, Bendle:2019csk, Chen:2022jux}. Evidently, the top-down approach of reduction is naturally related to the recursive structure of Baikov representations. In this work, we utilize this relationship to demonstrate how the recursive structure can be used to simplify the top-down reduction procedure.
	
	The contents are organized as follows. In Sec.~\ref{sec:recursive}, we briefly review the recursive structure of the Baikov representations presented in \cite{Jiang:2023qnl}. In Sec.~\ref{sec:algebraic}, we establish an algebraic framework for separating the system of Feynman integrals into disjoint subsystems using cuts. This provides a unified view on the different reduction approaches. In Sec.~\ref{sec:reduction} and Sec.~\ref{sec:reductionm}, we show how the recursive structure combined with the intersection theory can help us perform top-down reductions. We summarize in Sec.~\ref{sec:conclusion}.

	\section{A brief reminder of the recursive structure}\label{sec:recursive}
	
	In this section, we briefly introduce the Baikov representations and their recursive structure. For detailed derivations we refer the readers to \cite{Jiang:2023qnl}.
	
	A Feynman integral family consists of scalar integrals of the form
	\begin{equation} \label{eq:FI}
		I(a_1,a_2,\cdots,a_N;d)=\int\frac{\nd^{d}l_1\nd^{d}l_2\cdots\nd^{d}l_L}{(i\pi^{d/2})^L}\frac{1}{x_1^{a_1}x_2^{a_2}\cdots x_N^{a_N}} \, ,
	\end{equation}
	where $L$ is the number of loops and $d=4-2\epsilon$ is the dimension of spacetime; $N=L(L+1)/2+LE$ is the number of independent scalar products involving loop momenta, and $E$ is the number of independent external momenta (the number of external legs is thus $E+1$). The variables $x_{i}$ are propagator denominators if $a_i>0$ and irreducible scalar products (ISPs) if $a_{i}\le 0$. For the above integrals, we can write down the standard Baikov representation
	\begin{equation}\label{eq:baikovFI}
		I(a_1,\ldots,a_N;d)= C(p_1,\cdots,p_E;d) \int_{\mathcal{C}} \frac{\nd x_1\cdots\nd x_N}{x_1^{a_1}\cdots x_N^{a_N}} \left[ P^{L}_{N}(x_1,\cdots,x_N) \right]^{(d-K-1)/2} \, ,
	\end{equation}
	where $K=L+E$, and $C(p_1,\cdots,p_E;d)$ is an unimportant prefactor for our purpose, that will often be suppressed later. The integration contour ${\mathcal{C}}$ is determined by the polynomial
	\begin{equation}
		P^{L}_{N}(x_1,\ldots,x_N)=G(q_1,q_2,\ldots,q_K) \, ,
	\end{equation}
	where $\{q_1,q_2,\ldots,q_K\}$ denotes $\{l_1,\ldots,l_L,p_1,\ldots,p_E\}$, and $G$ represents the Gram determinant
	\begin{equation}\label{eq:detQ}
		G(q_1,q_2,\cdots,q_n) \equiv \det\left(\begin{array}{cccc} q_1^2 & q_1\cdot q_2 & \cdots & q_1\cdot q_n \\ q_2\cdot q_1 & q_2^2 & \cdots & q_2\cdot q_n \\ \vdots & \vdots & \ddots & \vdots \\ q_n\cdot q_1 & q_n\cdot q_2 & \cdots & q_n^2 \end{array}\right) .
	\end{equation}
	
	The standard Baikov representation works for all integrals within the family. However, for integrals in a given sub-sector, it is usually possible to integrate out some of the ISP variables. This leads to Baikov representations with fewer integration variables, which often coincide with the so-called loop-by-loop (LBL) representations. These representations (including the standard one) take the generic form
	\begin{equation}
		\label{eq:gen_rep}
		\int_{\mathcal{C}} \frac{\nd x_1\cdots\nd x_n}{x_1^{a_1}\cdots x_n^{a_n}} \left[ P_1(\bm{x}) \right]^{\gamma_1}  \cdots \left[ P_m(\bm{x}) \right]^{\gamma_m} \, ,
	\end{equation}
	where we use $\bm{x}$ to denote the sequence of variables $x_1,\ldots,x_n$ with $n \leq N$, and $P_1,\ldots,P_m$ are Baikov polynomials which are raised to non-integer powers $\gamma_1,\ldots,\gamma_m$. These various representations form a tree-like recursive structure starting from the standard one.
	
	An ISP variable to be integrated out must appear quadratically in one of the Baikov polynomials $P_j(\bm{x})$, but is absent in the other polynomials (we refer to it as a \emph{quadratic variable}). Denoting this variable as $z$, we may write $P_j(\bm{x})$ as
	\begin{equation}
		\label{eq:quad_poly}
		P_j(\bm{x}) = - (A z^2 + B z + C) = -A(z-c_1)(z-c_2) \, ,
	\end{equation}
	where $A$, $B$ and $C$ are polynomials of the remaining variables in $\bm{x}$. We can then integrate $z$ out using the \emph{recursion formula}
	\begin{multline}\label{eq:recursionformula}
		\int_{c_1}^{c_2}z^{n} \left[-A(z-c_1)(z-c_2)\right]^{\gamma}\mathrm{d}z=(-A)^{\gamma} \, (c_2-c_1)^{1+2\gamma} \, \frac{\Gamma(1+\gamma)^2}{\Gamma(2(1+\gamma))}
		\\
		\times \left(\frac{c_1+c_2}{2}\right)^{n} \, {}_{2}F_{1}\left(-\frac{n}{2},\frac{1-n}{2},\frac{3}{2}+\gamma;\left(\frac{c_1-c_2}{c_1+c_2}\right)^2\right) ,
	\end{multline}
	where $n \geq 0$. Note that the above hypergeometric function is actually always a polynomial of its last argument. The above procedure can be repeated for another quadratic variable if it exists, and arrive at representations with even fewer integration variables.

	\section{The top-down reduction and its algebraic structure}\label{sec:algebraic}
	
	Integral reduction is one of the bottlenecks in multi-loop multi-leg calculations. For cutting-edge problems it involves a large number of linear relations. One way to efficiently generate and solve these relations is the so-called ``top-down'' approach emerging naturally from generalized unitarity method and integrand reduction \cite{Britto:2004nc,Ossola:2006us,Kilgore:2007qr,Ellis:2008ir,Zhang:2012ce,Frellesvig:2020qot}. The idea is very simple. Given an integral to reduce, one first finds its top-sector components by solving the linear relations under the maximal cut. The maximal cut significantly reduces the number of variables and the number of equations, making the reduction much simpler. One then subtract the top-sector components from the integral, and transform the resulting integrand into sub-sectors (which is the most non-trivial part of this method). We can then employ the recursive structure of Baikov representations to arrive at lower representations, and repeat the above procedure by working with the maximal cut for the sub-sectors.
	
	In this Section, we briefly review the basic idea of the top-down reduction and the graded structure of the vector space of Feynman integrals. We will use the language of intersection theory \cite{Frellesvig:2019kgj, Frellesvig:2019uqt, Frellesvig:2020qot, Chestnov:2022xsy}, although in practice one may employ any method suitable for solving the linear system at hand.
	
	\subsection{The cohomological formulation of cuts for Feynman integrals}
	
	We first discuss the general algebraic structure of Feynman integrals, applicable within any representation where a Feynman integral takes the form
	\begin{equation}
	    \label{eq:uphi}
		I=\int_{\mathcal{C}} u \, \varphi \,,
	\end{equation}
	where $u$ is a multi-valued function on $\mathbb{C}^{N}$, $\mathcal{C}$ is an integration contour, and $\varphi$ is a single-valued $N$-form that will be referred to as an ``integrand''. The $u$-function vanishes on the boundary of $\mathcal{C}$, i.e., $u \big|_{\partial \mathcal{C}}=0$. The $N$-form $\varphi$ may have singularities on $\partial \mathcal{C}$, where the integral is regularized by the $u$-function. We call these regularized singular points as ``twisted boundaries''\cite{Mizera:2019ose}. In the same time, $\varphi$ may also have singularities at places other than $\{u=0\}$. We call these singular points as ``relative boundaries''. We will assume that the relative boundaries are given by the set $\{D=0\}\equiv\bigcup_{i=1}^{n}\{D_{i}=0\}$ with $n \leq N$, where the $D_i$'s are functions on $\mathbb{C}^{N}$ (which will be identified with propagator denominators). These relative boundaries are removed from the integration contour, and hence $\varphi$ is holomorphic within $\mathcal{C}$.
	
	Using Stoke's theorem, we have 
	\begin{equation}
		0=\int_{\mathcal{C}}d(u\xi)=\int_{\mathcal{C}}u\nabla_{\omega}\xi \,,
		\label{eq:ibpequiv}
	\end{equation}
	where $\nabla_{\omega}\equiv d+\omega \wedge$, $\omega\equiv d\log u$, and $\xi$ is a single-valued holomorphic $(N-1)$-form. The above equation generates integration-by-parts (IBP) identities among different integrals, which are used for integral reduction. Formally, the IBP equivalence can be encoded in the $N$th twisted cohomology group 
	\begin{equation}
		H^{N}(X;\nabla_{\omega})=\frac{\operatorname{ker}:\Omega^{N}(X)\to\Omega^{N+1}(X)}{\operatorname{im}:\Omega^{N-1}(X)\to\Omega^{N}(X)} \,,
	\end{equation}
	where $X\equiv \mathbb{C}^{N}\setminus{(\{u=0\}\cup\{D=0\})}$. An elements of $H^{N}$ is the equivalence class of integrands that give the same integral, $\bra{\varphi}: \varphi \sim \varphi + \nabla_{\omega}\xi$. The number of independent integrals, i.e., the dimension of $H^{N}$, is finite \cite{Smirnov:2010hn}. Moreover, in dimensional regularization,  $H^{N}$ is usually the only non-trivial cohomology group, as all other $H^{k \neq N}$ vanish \cite{aomoto1975vanishing}. Using the complex Morse theory, one can obtain $\nu = \dim(H^{N})$ by counting the number of critical points from the $u$-function \cite{Lee:2013hzt, Frellesvig:2019uqt}. One may then choose a basis $\bra{e_i}$ ($i=1,\ldots,\nu$), and decompose any integral $\bra{\varphi}$ as $\bra{\varphi} = \sum_{i=1}^\nu c_i \bra{e_i}$. This is just the IBP reduction procedure. To compute the coefficients $c_i$, one needs to solve a large linear system using, e.g., the intersection theory. In the intersection theory, one introduces the space of dual forms $\ket{\varphi} \in H^N(X;\nabla_{-\omega})$, and defines a pairing between dual forms and Feynman integrands called intersection numbers \cite{Mastrolia:2018uzb,Frellesvig:2019uqt,Frellesvig:2020qot}. The intersection number between $\bra{\varphi_L}$ and $\ket{\varphi_R}$ is given by
	\begin{equation}
	    \label{eq:intnum}
		\braket{\varphi_L | \varphi_R} \equiv \frac{1}{(2\pi i)^N} \int_X \varphi_L \wedge \varphi_R^c \,,
	\end{equation}
	where $\varphi_R^c$ is IBP equivalent to $\varphi_R$ but has compact support.
	
	To perform computations in the presence of relative boundaries $\{D=0\}$, one usually introduces additional regularizations to convert them into twisted boundaries. The simplest regularization is to multiply the $u$-function by factors such as $D_i^\rho$, where $\rho$ will be taken to zero in the end \cite{Mizera:2019ose, Frellesvig:2019uqt, Frellesvig:2020qot}. Although this technique is valid in many practical cases, there is no general proof that the $\rho \to 0$ limit is guaranteed to give the correct results. Recently, there arises a new technique based on the concept of twisted relative cohomology \cite{Caron-Huot:2021xqj, Caron-Huot:2021iev}. This bypasses the introduction of regularizations for relative boundaries, removing the ambiguities and extra efforts in the calculations.
	
	A key step in the works \cite{Caron-Huot:2021xqj, Caron-Huot:2021iev} is the decomposition of the full cohomology of dual integrands into subspaces corresponding to different cuts of Feynman integrands. Motivated by that, we suggest that one can achieve a similar decomposition of the cohomology of Feynman integrands, which provides a natural language to rigorously describe the top-down and bottom-up approaches for the reduction of Feynman integrals. In the top-down approach, the introduction of extra regularizations can also be avoided, which simplifies the computation significantly.
	
	To perform the decomposition, we utilize the following short exact sequence of Feynman integrands:\footnote{This can be obtained from a similar short exact sequence of dual integrands in \cite{Caron-Huot:2021xqj, Caron-Huot:2021iev}.} 
	\begin{equation}
	    \label{eq:short_sequence}
		0 \leftarrow H^{N-1}(D^{(i)}\setminus\{D_{\neq i}=0\}) \xleftarrow{\delta_{i}^{*}} H^{N}(Y\setminus \{D=0\})\xleftarrow{\iota_{i}^{*}} H^{N}(Y\setminus \{D_{\neq i}=0\}) \leftarrow 0 \,,
	\end{equation}
	where $Y\equiv\mathbb{C}^{N}\setminus\{u=0\}$, $D^{(i)}\equiv Y\cap\{D_{i}=0\}$, and $\{D_{\neq i}=0\} \equiv \bigcup_{j\neq i}\{D_{j}=0\}$. The full cohomology of Feynman integrands is just $H^{N}(Y\setminus \{D=0\})$, while $H^{N}(Y\setminus \{D_{\neq i}=0\})$ contains those integrands that have no singularity on $\{D_i = 0\}$. The map $\iota_{i}^{*}$ is the natural embedding. The cohomology $H^{N-1}(D^{(i)}\setminus\{D_{\neq i}=0\})$ contains $(N-1)$-forms which live inside $\{D_i = 0\}$, which correspond to the integrands after we impose cut on $D_i$. The map $\delta_{i}^{*}$ simply corresponds to this operation of cut.
	
	Given the short exact sequence, the following isomorphism holds:
	\begin{equation}
		H^{N}(Y\setminus \{D=0\}) \cong H^{N-1}(D^{(i)}\setminus\{D_{\neq i}=0\}) \oplus H^{N}(Y\setminus \{D_{\neq i}=0\}) \,.
		\label{eq:decomposition1}
	\end{equation}
	To discuss integral reduction, we need to look further into this isomorphism. While $H^{N}(Y\setminus \{D_{\neq i}=0\})$ is naturally embedded into $H^{N}(Y\setminus \{D=0\})$ by $\iota_{i}^{*}$, we need to define a map $\sigma_{i}$ as a pullback of $H^{N-1}(D^{(i)}\setminus\{D_{\neq i}=0\})$. This map acts as ``undoing a cut'', and allows us to rewrite the isomorphism as an identity:
	\begin{equation}
		H^{N}(Y\setminus \{D=0\}) = \sigma_{i} \left(H^{N-1}(D^{(i)}\setminus\{D_{\neq i}=0\})\right) \oplus \iota_{i}^{*} \left(H^{N}(Y\setminus \{D_{\neq i}=0\})\right).
		\label{eq:decomposition2}
	\end{equation}
	The pullback map $\sigma_{i}$ satisfies $\delta_{i}^{*}\circ \sigma_{i}=\operatorname{id}$ and is only unique modulo $\operatorname{ker}(\delta_{i}^{*})=\operatorname{im}(\iota_{i}^{*})$. Intuitively, the above decomposition is simply categorizing the integrands according to whether $D_i$ appears in the denominator.
	
	In principle, $\sigma_{i}$ can be constructed in any representation. However, its construction is particularly straightforward in the Baikov representations, where $D_i \equiv x_i$ themselves are integration variables. Hence, $\{D_i = 0\}$ is simply a coordinate hyperplane. Consider a $\bra{\psi} \in H^{N-1}(D^{(i)}\setminus\{D_{\neq i}=0\})$ given by 
	\begin{equation}
		\psi = \hat{\psi} \, dx_{1}\wedge\cdots\wedge\widehat{dx_{i}}\wedge\cdots\wedge dx_{N} \,,
	\end{equation}
	where $\hat{\psi}$ is a function, and $\widehat{dx_{i}}$ means that this factor is absent. We can naturally assign $\sigma_i(\bra{\psi}) = \bra{\varphi}$ where
	\begin{equation}
		\varphi = \frac{\hat{\psi}}{x_{i}} \, dx_{1}\wedge\cdots\wedge dx_{i} \wedge\cdots\wedge dx_{N} \,.
	\end{equation}
	
	The above procedure can be repeated to decompose $H^{N-1}(D^{(i)}\setminus\{D_{\neq i}=0\})$. For that we choose the next relative boundary $\{D_j = 0\}$. The short exact sequence is then given by
	\begin{align}
		0 \leftarrow H^{N-2}(D^{(i,j)}\setminus\{D_{\neq i,j}=0\}) &\leftarrow H^{N-1}(D^{(i)}\setminus\{D_{\neq i}=0\}) \nonumber \\
		&\leftarrow H^{N-1}(D^{(i)}\setminus\{D_{\neq i,j}=0\}) \leftarrow 0 \,,
	\end{align}
	where $D^{(i,j)} \equiv D^{(i)} \cap D^{(j)}$. This tells us that
	\begin{equation}
		H^{N-1}(D^{(i)}\setminus\{D_{\neq i}=0\}) \cong H^{N-2}(D^{(i,j)}\setminus\{D_{\neq i,j}=0\}) \oplus H^{N-1}(D^{(i)}\setminus\{D_{\neq i,j}=0\}) \,.
	\end{equation}
	Similarly, the second term in Eq.~\eqref{eq:decomposition1} can be decomposed as
	\begin{equation}
		H^{N}(Y\setminus \{D_{\neq i}=0\}) \cong H^{N-1}(D^{(j)}\setminus\{D_{\neq i,j}=0\}) \oplus H^{N}(Y\setminus\{D_{\neq i,j}=0\}) \,.
	\end{equation}
	Recursively applying the decomposition, we can finally arrive at
	\begin{equation}
		\label{eq:fulldirectsum}
		H^{N}(Y\setminus\{D=0\}) \cong \bigoplus_{I\subseteq\{1,\ldots,n\}} H^{N-|I|}(D^{(I)}) \,,
	\end{equation}
	where $|I|$ denotes the cardinality of the subset $I$, $D^{(I)} \equiv \bigcap_{i \in I} D^{(i)}$, and $D^{(\varnothing)}\equiv Y$. Intuitively, $H^{N-|I|}(D^{(I)})$ is the cohomology of integrals that survive after cutting all $D_i$ for $i \in I$, and vanish when cutting any further propagators. In other words, it is ``the top-sector of a sub-sector''. Note that after the decomposition, there are no relative boundaries in each component. Hence, no additional regularization is required to perform the calculation. This provides the algebraic framework of the top-down reduction approach.
	
	Note that the direct-sum decomposition of the full space can be used to count the dimension (i.e., the number of master integrals) by adding together the dimensions of the subspaces. Namely,
	\begin{equation}
		\dim(H^{N}(Y\setminus\{D=0\})) = \sum_{I\subseteq\{1,\ldots,n\}} \dim(H^{N-|I|}(D^{(I)})) \,.
	\end{equation}
	When computing each $\dim(H^{N-|I|}(D^{(I)}))$, there is no need to introduce extra regulators. By the same reasoning, we can consider the cohomology corresponding to a sub-sector:
	\begin{equation}
		H^{N-|I|}(D^{(I)}\setminus\{D_{\notin I}=0\}) \cong \bigoplus_{J\supseteq I}H^{N-|J|}(D^{(J)}) \,,
	\end{equation}
	and count the number of master integrals in a similar way. Hence, the decomposition provides us a general and rigorous way to calculate the dimension of the space of integrals at any level between the maximal cut and the full family (see relevant discussions in \cite{Mizera:2019ose}). In the literature, an alternative way to count the number of master integrals in a sub-sector is to work in the full space with regulators applied for the propagator denominators in that sub-sector. It is not entirely clear whether it always yield correct results. It is worthwhile to investigate further the relationship between the two counting methods.
	
	\subsection{Reduction approaches in the cohomological language}
	
	We now move to discuss the different reduction approaches in the cohomological language in the previous subsection. In particular, we consider the three approaches to decompose Feynman integrals by intersection theory described in \cite{Frellesvig:2020qot}: straight, bottom-up and top-down.
	
	The straight decomposition is conceptually the simplest. One just directly computes intersection numbers in the space $H^{N}(Y\setminus \{D=0\})$, i.e., without imposing any cut. Due to the existence of relative boundaries, one must introduce regulators for all propagator denominators.
	
	The bottom-up decomposition proceeds by choosing a list of spanning cuts. Each cut in the list corresponds to an $I_i \subset \{1,\ldots,n\}$, and one computes intersection numbers in the subspace
	\begin{equation}
		H^{N-|I_{i}|}(D^{(I_{i})}\setminus\{D_{\notin I_{i}}=0\}) \cong \bigoplus_{J \supseteq I_i} H^{N-|J|}(D^{(J)}) \,,
	\end{equation}
	which contains all integrands that survive the cut. The calculation within this subspace is simpler than in the full space, because fewer integration variables are involved. Nevertheless, one still needs to introduce regularizations for the remaining propagator denominators $D_{\notin I_{i}}$.
	
	Finally, we discuss the top-down reduction approach in some detail. Suppose that we want to reduce $\bra{\varphi} \in H^{N}(Y\setminus\{D=0\})$ as a linear combination of a basis of $H^{N}(Y\setminus\{D=0\})$. According to Eq.~\eqref{eq:fulldirectsum}, we know that it can be decomposed as a sum of components belonging to each of the subspace in the direct sum. In particular, there is a component $\bra{\varphi_{M}} \in H^{N-n}(D^{(1,\cdots,n)})$. This is simply the top-sector component under the maximal cut. Within $H^{N-n}(D^{(1,\cdots,n)})$, we can compute the intersection numbers over $N-n$ variables, which is in practice very easy since $N-n$ is usually a small number. We also emphasize again that one does not need to introduce extra regularizations here due to the absence of relative boundaries (there are in fact no propagators remaining). We assume that by computing the intersection numbers, $\bra{\varphi_{M}}$ can be reduced as
	\begin{equation}
	    \label{eq:phi_M}
		\bra{\varphi_{M}} = c_{1} \bra{e_{1}} + \cdots + c_{\nu} \bra{e_{\nu}} \,,
	\end{equation}
	where $\nu=\dim(H^{N-n}(D^{(1,\cdots,n)}))$ and $\{\bra{e_i}\}$ is a basis of $H^{N-n}(D^{(1,\cdots,n)})$. We may then pullback $\bra{\varphi_{M}}$ to the original space and subtract it from $\bra{\varphi}$:
	\begin{equation}
	    \label{eq:phi_r}
		\bra{\varphi_{r}} \equiv \bra{\varphi} - \sigma(\bra{\varphi_{M}}) = \bra{\varphi} - c_{1}\sigma(\bra{e_{1}})-\cdots-c_{k}\sigma(\bra{e_{k}}) \,,
	\end{equation}
	where $\sigma$ is the composition of all $n$ pullback maps $\sigma_i$ when going through the decompositions as in Eq.~\eqref{eq:decomposition2}. For notational convenience, we will often make the $\sigma$-maps implicit in the following.
	
	Now we know that $\bra{\varphi_{r}}$ must belong to the subspace where $H^{N-n}(D^{(1,\cdots,n)})$ is removed from the direct sum \eqref{eq:fulldirectsum}, i.e.,
	\begin{equation}
		\bra{\varphi_{r}} \in \bigoplus_{I\subset \{1,\ldots,n\}} H^{N-|I|}(D^{(I)}) \,,
	\end{equation}
	where the subspaces $H^{N-|I|}(D^{(I)})$ are implicitly pulled-back or embedded into the full space through the $\sigma$- and $\iota^*$-maps. However, it is highly nontrivial to find a representative $\varphi_r$ that \emph{explicitly} takes the form of subspace integrands. Nevertheless, from the above we know that the existence of such a representative is guaranteed. In later Sections, we demonstrate how to systematically find it via intersection theory.
	
	Given a suitable representative $\varphi_r$, the next step is to apply the procedure recursively. We know that $\bra{\varphi_{r}}$ may have components in the following $n$ subspaces:
	\begin{equation}
		H^{N-n+1}(D^{(2,\cdots,n)}) ,\, H^{N-n+1}(D^{(1,3,\cdots,n)}),\, \ldots ,\, H^{N-n+1}(D^{(1,\cdots, n-1)}) \,.
	\end{equation}
	In each subspace, we again apply maximal cut and repeat the above procedure, until we arrive at the lowest sub-sectors. We emphasize that in the whole process of computation, there is no relative boundary involved and hence no need for regularization.
	
	At first sight, it seems that for a sub-sector with $m$ propagators, we need to compute intersection numbers over $N-m$ variables (see, e.g., examples in \cite{Frellesvig:2020qot}). In lower and lower sub-sectors, this becomes more and more complicated. This is where the recursive structure of Baikov representations comes into play. In sub-sectors, we can employ the recursion formula \eqref{eq:recursionformula} to integrate out some ISPs. As a result, we arrive at a representation with $N' < N$ variables. We then only need to compute intersection numbers over $N'-m$ variables, which is in practice very easy. In particular, for one-loop reductions we actually don't need to compute any intersection number. This is in contrast to the top-down approach outlined in \cite{Frellesvig:2020qot}, and shows the advantage of our approach.

	\section{Integrand reduction for the top-down approach}\label{sec:reduction}
	
	From the discussions in the previous Section, we see that the top-down approach avoids the regularization of relative boundaries, and significantly reduces the complexity of the intersection numbers required for the reduction. However, a key step in this approach is to transform the top-sector subtracted integrand (i.e., $\bra{\varphi_r}$ in Eq.~\eqref{eq:phi_r}) into a form that explicitly belongs to sub-sectors. In \cite{Frellesvig:2020qot}, this is done by introducing an ansatz that takes the desired form, with coefficients to be determined by IBP relations. In this Section, we provide a systematic method to achieve that transformation, and demonstrate our method with several examples.	
	
	\subsection{Top-sector ISP reduction in the Baikov representations}\label{subsec:fullreduction}
	
	From Eq.~\eqref{eq:gen_rep}, we see that the Baikov representations take exactly the form of Eq.~\eqref{eq:uphi} studied in the previous Section. In particular, the relative boundaries $D_i=0$ simply correspond to the vanishing surfaces of the propagator denominators. To setup the notation, we will focus on one sector with propagator denominators $\bm{x}_{\prop} = \{x_1,\ldots,x_n\}$, and ISPs $\bm{x}_{\isp} = \{x_{n+1},\ldots,x_N\}$. We also use $\bm{x}$ to denote the union of $\bm{x}_{\prop}$ and $\bm{x}_{\isp}$. The integrals in this sector (including sub-sectors) can then be written as
	\begin{align}\label{eq:generalform}
		I(a_1,...,a_{N}) = \int_{\mathcal{C}} u \, \varphi = \int_{\mathcal{C}} \mathrm{d}^n \bm{x}_{\prop} \, \mathrm{d}^{m} \bm{x}_{\isp} \, u(\bm{x}) \prod_{i=1}^{N} x_{i}^{-a_{i}} \, ,
	\end{align}	
	where $m=N-n$, and we have suppressed the unimportant pre-factors.
	
	Consider an integrand $\varphi$ in the top sector, i.e., with all $a_1,\ldots,a_n$ being positive. Recall from Eq.~\eqref{eq:phi_M} that the first step in top-down reduction is to decompose the integral $\bra{\varphi}$ as a linear combination of the top-sector masters up to sub-sector components. Here, we use a slightly different notation:
	\begin{equation}
	    \label{eq:phi_top_decompose}
		\bra{\varphi} = c_{1} \bra{e_{1}} + \cdots + c_{\nu} \bra{e_{\nu}} + \text{sub-sector integrals} \,,
	\end{equation}
	where $\{\bra{e_{1}},\ldots,\bra{e_{\nu}}\}$ is the pullback of a basis of the subspace $H^{N-n}(D^{(1,\cdots,n)})$. In other words, let $\bra{e_{M,i}} = \delta^*\bra{e_{i}}$, where $\delta^*$ is the composition of the mappings $\delta^*_i$ defined in Eq.~\eqref{eq:short_sequence}, which simply corresponds to the maximal cut. Then $\{\bra{e_{M,1}},\ldots,\bra{e_{M,\nu}}\}$ is a basis of $H^{N-n}(D^{(1,\cdots,n)})$, 
	
	The coefficients $c_{i}$ can be computed as intersection numbers in the full space $H^N$, i.e., without any cut. Nevertheless, it is much simpler to work in the subspace $H^{N-n}(D^{(1,\cdots,n)})$, which is the main benefit of the top-down approach. We can compute the intersection numbers under the maximal cut, i.e., taking the residues at the origin with respect to all propagator denominators, and only integrating over the ISPs in the formula \eqref{eq:intnum}. To be more precise, for the dual space of $H^{N-n}(D^{(1,\cdots,n)})$, we find a dual basis $\{\ket{d_{M,1}},\ldots,\ket{d_{M,\nu}}\}$ which satisfies $\braket{e_{M,i} | d_{M,j}} = \delta_{ij}$. The coefficients are then given by $c_i = \braket{\varphi_M | d_{M,i}}$, where $\bra{\varphi_M} = \delta^*\bra{\varphi}$.

	After obtaining the coefficients, we go back to the full space $H^N$. We define the top-sector subtracted integrand of $\varphi$ as
	\begin{equation}
	    \label{eq:subtracted}
	    \varphi_r = \varphi - \sum_{i=1}^{\nu} c_i \, e_i \,,
	\end{equation}
    where $e_i$ is an arbitrary representative of $\bra{e_i}$. In general, $\varphi_r$ may still contain top-sector terms with all $a_1,\ldots,a_n$ being positive in the integrand level, although $\bra{\varphi_r}$ has no top-sector component in the integral level. We are now going to discuss how to bring $\varphi_r$ into a form that explicitly has no top-sector term. Before that, we first introduce two useful concepts.
    
	The first concept is the \emph{regular form} of Feynman integrals in Baikov representations. For a given integral $I(a_1,\ldots,a_N)$, its integrand can be written in many equivalent ways. In particular, we can use IBP transformation to make the powers of all propagators to be $1$ in the denominator, at the price of higher power terms of ISPs in the numerator. Defining the partial differentiation operator
    \begin{equation}
		\mathcal{D}^{\vec{a}} = \frac{1}{\prod_{i=1}^{n} \Gamma(a_i)} \left(\frac{\partial}{\partial x_1}\right)^{a_{1}-1} \cdots \left(\frac{\partial}{\partial x_n}\right)^{a_n-1} ,
	\end{equation}
    with $\vec{a}=(a_1,\ldots,a_n)$, we can write the regular form of an integral as
	\begin{equation}\label{eq:regularform}
		I(a_1,\ldots,a_N) = \int_{\mathcal{C}} \mathrm{d}^n \bm{x}_{\prop} \, \mathrm{d}^{m} \bm{x}_{\isp} \, \frac{u(\bm{x})}{x_1 \cdots x_n} \left( \prod_{j=n+1}^{N} x_{j}^{-a_{j}} \right) \frac{\mathcal{D}^{\vec{a}} u(\bm{x})}{u(\bm{x})} \,.
	\end{equation}	
	It is convenient to perform the maximal cut in this form, which boils down to the replacement $1/x_{i}\to \delta(x_{i})$ for $i=1,\ldots,n$. Note that the $u$ function takes the form
	\begin{equation}
	    u(\bm{x}) =  \left[ P_1(\bm{x}) \right]^{\gamma_1}  \cdots \left[ P_m(\bm{x}) \right]^{\gamma_m} \,,
	\end{equation}
	with non-integer powers $\gamma_1,\ldots,\gamma_m$. Hence, the combination $\mathcal{D}^{\vec{a}} u(\bm{x}) / u(\bm{x})$ generally contains polynomials $P_1,\ldots,P_m$ in the denominator. As a result, Eq.~\eqref{eq:regularform} belongs to the so-called \emph{generalized Baikov representions} \cite{Chen:2022lzr}. Nevertheless, there is no problem to study their IBP relations using intersection theory.
	
	The second concept is the \emph{top-sector ISP reduction} of an integral within the intersection theory. We note that the coefficients in Eq.~\eqref{eq:phi_top_decompose} are computed under the maximal cut. That is, we take the residues of $e_1,\ldots,e_\nu$ and $\varphi$ at $\bm{x}_{\prop} = \bm{0}$, before doing the computations. We now propose to compute a different set of intersection numbers, where $\bm{x}_{\prop}$ are regarded as external parameters instead of being taken to zero. Precisely speaking, we are considering the IBP relations among ISP-integrated partial integrals in the regular form
	\begin{equation}
	\label{eq:partial_integrals}
	\tilde{I}(a_1,\ldots,a_N) \equiv \int_{\tilde{\mathcal{C}}} u \, \tilde{\varphi} = \int_{\tilde{\mathcal{C}}} \mathrm{d}^{m} \bm{x}_{\isp} \, \frac{u(\bm{x})}{x_1 \cdots x_n} \left( \prod_{j=n+1}^{N} x_{j}^{-a_{j}} \right) \frac{\mathcal{D}^{\vec{a}} u(\bm{x})}{u(\bm{x})} \,.
	\end{equation}
	The original integrals defined in Eq.~\eqref{eq:generalform} can be obtained by further integrating over $\bm{x}_{\prop}$ from the above. 
	
	The equivalence classes of integrals of the above form belongs to a cohomology group $H^{N-n}(\mathbb{C}^{N-n}\setminus\{u=0\}; \bm{x}_{\prop})$, where $\bm{x}_{\prop}$ are regarded as external parameters. It should be noted that the dimension $\tilde{\nu}$ of this space is not necessarily the same as the dimension $\nu$ of the maximal cut space $H^{N-n}(D^{(1,\cdots,n)})$. Assuming that a basis is given by $\{\bra{\tilde{e}_1},\ldots,\bra{\tilde{e}_{\tilde{\nu}}}\}$ (which should always contain the $x_1 \cdots x_n$ factor in the denominator), we can decompose any element $\bra{\tilde{\varphi}}$ as:
	\begin{equation}\label{eq:fullreduction}
		\bra{\tilde{\varphi}} = \sum_{i=1}^{\tilde{\nu}} \tilde{c}_i(\bm{x}_{\prop}) \bra{\tilde{e}_{i}} \, ,
	\end{equation}
	which is defined as the \emph{top-sector ISP reduction}. The coefficients $\tilde{c}_i(\bm{x}_{\prop})$ can again be computed as $N-n$ variable intersection numbers. We can expect that in the limit $\bm{x}_{\prop} \to \bm{0}$, the set of coefficients $\{\tilde{c}_1,\ldots,\tilde{c}_{\tilde{\nu}}\}$ must be related to the set of coefficients $\{c_1,\ldots,c_{\nu}\}$ in Eq.~\eqref{eq:phi_top_decompose}. One possible complication is that $\tilde{\nu}$ can be larger than $\nu$. However, in this case it happens that in the limit $\bm{x}_{\prop} \to \bm{0}$, some of the integrals in $\{\bra{\tilde{e}_1},\ldots,\bra{\tilde{e}_{\tilde{\nu}}}\}$ become reducible, and $\{\tilde{c}_1,\ldots,\tilde{c}_{\tilde{\nu}}\}$ indeed become $\{c_1,\ldots,c_{\nu}\}$ after appropriate recombinations.
	The simplest method to account for this is to \emph{define} $e_i$ from $\tilde{e}_i$, and $c_i$ from $\tilde{c}_i$. In this way, there will be redundant integrals in the set $\{\bra{e_1},\ldots,\bra{e_{\tilde{\nu}}}\}$, which can be taken care of later by a further reduction.
	Hence, in the following we will assume $\nu = \tilde{\nu}$ and
	\begin{equation}\label{eq:crelation}
		c_{i} = \tilde{c}_{i}(\bm{x}_{\prop}=\bm{0}) \, , \quad (i=1,\ldots,\nu) \, .
	\end{equation}
	Here, we have implicitly assumed that the limit $\tilde{c}_i(\bm{x}_{\prop} \to \bm{0})$ exists. We believe that this is always the case whenever $\nu = \tilde{\nu}$. When $\tilde{\nu} > \nu$, on the other hand, some of the $\tilde{c}_i$ coefficients may become singular in that limit. This is not a problem since, after the recombinations mentioned above, these singular behaviors will disappear. We will encounter such situations in Sec.~\ref{sec:reductionm}.
		
	We now want to study the top-sector subtracted integrand \eqref{eq:subtracted}. For simplicity, we first deal with the situations where $a_1=\cdots=a_n=1$ in $\varphi$, i.e., all powers of propagators are unity:
	\begin{equation}
		\varphi = \hat{\varphi} \, \mathrm{d}^n \bm{x}_{\prop} \, \mathrm{d}^{m} \bm{x}_{\isp} = \frac{Q(\bm{x}_{\isp})}{x_1 \cdots x_{n}} \, \mathrm{d}^n \bm{x}_{\prop} \, \mathrm{d}^{m} \bm{x}_{\isp} \, ,
	\end{equation}
	where $Q$ is a polynomial of $\bm{x}_{\isp}$. The integrands for a top-sector basis can be chosen as
	\begin{equation}
	    \label{eq:top_sector_basis}
		e_i = \hat{e}_i \, \mathrm{d}^n \bm{x}_{\prop} \, \mathrm{d}^{m} \bm{x}_{\isp} = \frac{Q_i(\bm{x}_{\isp})}{x_1 \cdots x_{n}} \, \mathrm{d}^n \bm{x}_{\prop} \, \mathrm{d}^{m} \bm{x}_{\isp} \, , \quad (i=1,\ldots,\nu) \,.
	\end{equation}
	We introduce a constant factor $D_0$ (hereafter, ``constant'' means only depending on $\epsilon$ and external momenta), and define
	\begin{equation}
	    \label{eq:N0}
	    N_0(\bm{x}_{\isp}) = D_0 \sum_{i=1}^{\nu} c_i \, Q_i(\bm{x}_{\isp}) \, .
	\end{equation}
	We can then write the top-sector subtracted integrand as $\varphi_r = \hat{\varphi}_r \, \mathrm{d}^n \bm{x}_{\prop} \, \mathrm{d}^{m} \bm{x}_{\isp}$, with
	\begin{equation}\label{eq:remainingintegrand}
			\hat{\varphi}_{r}=\hat{\varphi} - \sum_{i=1}^{\nu}c_{i} \, \hat{e}_{i}= \frac{Q(\bm{x}_{\isp})}{x_1 \cdots x_{n}} - \frac{N_{0}(\bm{x}_{\isp})}{D_{0} \, x_1 \cdots x_{n}}
			=\frac{Q(\bm{x}_{\isp}) \, D_{0} - N_{0}(\bm{x}_{\isp})}{D_{0} \, x_1 \cdots x_{n}} \, .
	\end{equation}
	The expression of $D_0$ is arbitrary at this point since it will be cancelled in the expression. It can be chosen for the convenience of calculation, as we'll see later.
	
	We now note that from the same $\hat{\varphi}$ and $\{\hat{e}_i\}$, we can define integrands for Eq.~\eqref{eq:partial_integrals}:
	\begin{equation}
	    \tilde{\varphi} = \hat{\varphi} \, \mathrm{d}^{m} \bm{x}_{\isp} \, , \quad \tilde{e}_i = \hat{e}_i \, \mathrm{d}^{m} \bm{x}_{\isp} \, .
	\end{equation}
	From the top-sector ISP reduction \eqref{eq:fullreduction}, we know that the following subtracted integrand gives vanishing result after integration over $\bm{x}_{\isp}$:
	\begin{equation}
	    \label{eq:isp_subtracted}
		\hat{\varphi} - \sum_{i=1}^{\nu} \tilde{c}_{i}(\bm{x}_{\prop}) \, \hat{e}_{i}
		=\frac{Q(\bm{x}_{\isp})D_{1}(\bm{x}_{\prop})-N_{1}(\bm{x}_{\isp},\bm{x}_{\prop})}{D_{1}(\bm{x}_{\prop}) \, x_1 \cdots x_{n}} \, ,
	\end{equation}
	where we have again introduced a polynomial factor $D_1(\bm{x}_{\prop})$, which may also depend on $\epsilon$ and external  momenta. The numerator $N_1(\bm{x})$ is defined as
	\begin{equation}
	    \label{eq:N1}
	    N_1(\bm{x}_{\isp},\bm{x}_{\prop}) = D_1(\bm{x}_{\prop}) \sum_{i=1}^{\nu} \tilde{c}_i(\bm{x}_{\prop}) \, Q_i(\bm{x}_{\isp}) \, .
	\end{equation}
	The vanishing of the integral is not affected if we rescale Eq.~\eqref{eq:isp_subtracted} by any factor independent of $\bm{x}_{\isp}$ (but maybe dependent on $\bm{x}_{\prop}$). Hence,
	\begin{equation}
		0=\int_{\tilde{\mathcal{C}}} \mathrm{d}^{m} \bm{x}_{\isp} \, u(\bm{x}) \, C_{0} \, \frac{Q(\bm{x}_{\isp})D_{1}(\bm{x}_{\prop})-N_{1}(\bm{x}_{\isp},\bm{x}_{\prop})}{D_{0} \, x_1 \cdots x_{n}} \, ,
	\end{equation}
	where we have replaced $D_{1}(\bm{x}_{\prop})$ in the denominator with $D_{0}$, and $C_0$ is another constant factor to be determined later.
		
	Now, we can subtract the above integrand from $\hat{\varphi}_r$ without altering the outcome of the integral, i.e.,
	\begin{multline}\label{eq:remainingintsub}
		I_{r} = \int_{\mathcal{C}} u \, \varphi_r = \int_{\mathcal{C}} \mathrm{d}^n \bm{x}_{\prop} \, \mathrm{d}^{m} \bm{x}_{\isp} \, u(\bm{x})
		\\
		\times \frac{Q(\bm{x}_{\isp})[D_{0}-C_{0} \, D_{1}(\bm{x}_{\prop})] - [N_{0}(\bm{x}_{\isp})-C_{0} \, N_{1}(\bm{x}_{\isp},\bm{x}_{\prop})]}{D_{0} \, x_1 \cdots x_{n}} \, .
	\end{multline}
	On the other hand, we can deduce from Eqs.~\eqref{eq:crelation}, \eqref{eq:N0} and \eqref{eq:N1} that
	\begin{equation}
	    \label{eq:C0}
		\frac{N_{0}(\bm{x}_{\isp})}{N_{1}(\bm{x}_{\isp},\bm{x}_{\prop}=\bm{0})} = \frac{D_{0}}{D_{1}(\bm{x}_{\prop}=\bm{0})} \, .
	\end{equation}
	Choosing $C_0$ to be the above ratio, we find that the numerator of \eqref{eq:remainingintsub} vanishes when $\bm{x}_{\prop} \to \bm{0}$.	Since this numerator is a polynomial of $\bm{x}_{\prop}$, it follows that each term of it must be proportional to some $x_i$ in $\bm{x}_{\prop}$. This will cancel the factor of $x_i$ in the denominator, leading to an integrand belonging to sub-sectors.
	
	In the above, all propagators in $\varphi$ are chosen to be power $1$ in $\varphi$. We now perform a similar analysis for a general $\varphi$ with powers of propagators being $\vec{a}=(a_1,\ldots,a_n)$. In the regular form, it can be written as
	\begin{equation}
	    \label{eq:phi_with_dots}
		\hat{\varphi}= \frac{Q(\bm{x}_{\isp}) \, \mathcal{D}^{\vec{a}}u(\bm{x})}{x_1 \cdots x_{n} \, u(\bm{x})} \, .
	\end{equation}
	Taking the top-sector basis as in Eq.~\eqref{eq:top_sector_basis}, we can write the top-sector subtracted integrand as
	\begin{equation}
			\hat{\varphi}_{r}=\hat{\varphi} - \sum_{i=1}^{\nu}c_{i} \, \hat{e}_{i}
			=\frac{Q(\bm{x}_{\isp}) \, D_{0} \, \mathcal{D}^{\vec{a}}u(\bm{x}) - N_{0}(\bm{x}_{\isp}) \, u(\bm{x})}{D_{0} \, x_1 \cdots x_{n} \, u(\bm{x})} \, .
	\end{equation}
	where the definitions of $N_{0}(\bm{x}_{\isp})$ and $D_{0}$ are the same as in \eqref{eq:remainingintegrand}.
	By exploiting the top-sector ISP reduction in the same way, we can transform the subtracted integrand into the form
	\begin{align}
		I_{r}&=\int_{\mathcal{C}} u \, \varphi_{r}=I_{r}^{(0)}+I_{r}^{(1)} \,, \nonumber \\
		I_{r}^{(0)}&=\int_{\mathcal{C}} \mathrm{d}^n \bm{x}_{\prop} \, \mathrm{d}^{m} \bm{x}_{\isp} \, \frac{(\mathcal{D}^{\vec{a}}u(\bm{x})) \, Q(\bm{x}_{\isp}) \left[D_{0}-C_{0}\,D_{1}(\bm{x}_{\prop})\right]}{D_{0} \, x_1 \cdots x_n} \nonumber \\
		&=(-1)^{|a|-n} \int_{\mathcal{C}} \mathrm{d}^n \bm{x}_{\prop} \, \mathrm{d}^{m} \bm{x}_{\isp} \, u(\bm{x}) \, Q(\bm{x}_{\isp}) \, \mathcal{D}^{\vec{a}}\frac{D_{0}-C_{0}D_{1}(\bm{x}_{\prop})}{D_{0} \, x_1 \cdots x_n} \,, \nonumber \\
		I_{r}^{(1)}&=-\int_{\mathcal{C}} \mathrm{d}^n \bm{x}_{\prop} \, \mathrm{d}^{m} \bm{x}_{\isp} \, u(\bm{x}) \, \frac{N_{0}(\bm{x}_{\isp})-C_{0}N_{1}(\bm{x}_{\isp},\bm{x}_{\prop})}{D_{0} \, x_1 \cdots x_n} \,,
	\end{align}
	where $|a|=\sum_{i=1}^{n} a_{i}$. Here we note that for $I^{(0)}_{r}$, we have applied integration-by-parts to move the derivatives from $u(\bm{x})$ to the rest of the integrand. This guarantees that the integrand explicitly corresponds to a Feynman integral, i.e., without polynomials in the denominator. Setting $C_{0}$ as in \eqref{eq:C0}, we again find that some of the $x_i$'s denominator for $1 \leq i \leq n$ must be cancelled by factors in the numerator, and the integrand degenerates to sub-sectors.
	
	Finally, it is also possible to choose master integrals with higher powers of propagators. Let's assume
	\begin{equation}
		\hat{e}_{i}= \frac{Q_{i}(\bm{x}_{\isp}) \, \mathcal{D}^{\vec{b}^{(i)}}u(\bm{x})}{x_1 \cdots x_{n} \, u(\bm{x})}  \,,	
	\end{equation} 
	where $\vec{b}^{(i)}=(b^{(i)}_{1},\ldots,b^{(i)}_{n})$. The top-sector subtracted integrand can then be written as
	\begin{align}\label{eq:generalreduction}
		I_{r}&=I^{(0)}_{r}+\sum_{i}I^{(i)}_{r} \nonumber
		\\
		I_{r}^{(0)}&=(-1)^{|a|-n} \int_{\mathcal{C}} \mathrm{d}^n \bm{x}_{\prop} \, \mathrm{d}^{m} \bm{x}_{\isp} \, u(\bm{x}) \, Q(\bm{x}_{\isp}) \, \mathcal{D}^{\vec{a}}\frac{D_{0}-C_{0}D_{1}(\bm{x}_{\prop})}{D_{0} \, x_1 \cdots x_n} \,, \nonumber
		\\
		I_{r}^{(i)}&=(-1)^{|b^{(i)}|-n+1} \int_{\mathcal{C}} \mathrm{d}^n \bm{x}_{\prop} \, \mathrm{d}^{m} \bm{x}_{\isp} \, u(\bm{x}) \, Q_i(\bm{x}_{\isp}) \, \mathcal{D}^{\vec{b}^{(i)}} \frac{N^{(i)}_{0}-C_{0}N^{(i)}_{1}(\bm{x}_{\prop})}{D_{0} \, x_1 \cdots x_n} \,,
	\end{align}
	where $|b^{(i)}|=\sum_{j=1}^{n}b^{(i)}_{j}$, and
	\begin{align}
	    N_0^{(i)} = c_i \, D_0 \, , \quad N_1^{(i)}(\bm{x}_{\prop}) = \tilde{c}_i(\bm{x}_{\prop}) D_1(\bm{x}_{\prop}) \,.
	\end{align}
	The constant $C_0$ is then chosen as
	\begin{equation}
	    \label{eq:generalreduction_C0}
		C_0 = \frac{D_0}{D_{1}(\bm{x}_{\prop}=0)}=\frac{N_{0}^{(i)}}{N_{1}^{(i)}(\bm{x}_{\prop}=0)} \,, \quad (i=1,\ldots,\nu) \,,
	\end{equation}
	and everything follows.

	\subsection{General one-loop reduction}

	As the first application of the formalism, we consider the one-loop case, which is simple but illustrating. At one-loop, all ISPs (inherited from super-sector representations) can be integrated out. Therefore, for each sector, we can always choose a representation with no ISPs. This representation involves only one Baikov polynomial $P(\bm{x})=P(\bm{x}_{\prop})$. The $u$-function is simply given by $P(\bm{x})^\gamma$, where $\gamma=(d-n-1)/2$ with $n$ being the number of propagators. In each sector, there is at most one master integral.
	
	We first discuss the reducible sectors, which do not have a master. This happens when $P(\bm{x}=\bm{0}) = 0$, i.e., the Baikov polynomial vanishes under maximal cut. It is easy to transform integrands in this sector to sub-sectors by dimensional recurrence relations \cite{Laporta:2000dc,Laporta:2000dsw,Lee:2009dh}. Roughly speaking, we write
	\begin{equation}
	    \label{eq:one_loop_dim_shift}
		P(\bm{x})^{\gamma} \frac{1}{x_1 \cdots x_n} = P(\bm{x})^{\gamma-1} \frac{P(\bm{x})}{x_1 \cdots x_n} \,.
	\end{equation}
	Since there is no constant term in the polynomial $P(\bm{x})$, each term in the numerator must cancel some propagator in the denominator. This leads to sub-sector integrals in a shifted dimension, which can then be brought back to $4-2\epsilon$ dimensions via dimensional recurrence relations. We refer the readers to Appendix~\ref{app:dimensionshift} for details. The dimension shift can also be performed using \texttt{LiteRed} \cite{Lee:2013mka}.
	
	We now discuss the normal sectors where $P(\bm{0})\neq 0$. Consider an integrand in the regular form
	\begin{equation}
		\hat{\varphi} = \frac{\mathcal{D}^{\vec{a}}P(\bm{x})^{\gamma}}{x_1 \cdots x_n \, P(\bm{x})^{\gamma}} \equiv \frac{N(\bm{x})}{x_1 \cdots x_n \, D(\bm{x})} \, ,
	\end{equation}
	where $D(\bm{x})=P(\bm{x})^{|\vec{a}|}$ and $N(x)$ is the corresponding numerator after cancelling the common factors of $P(\bm{x})$. Choosing the master integral in this sector as $\hat{e}_{1}=1/\left(x_{1} \cdots x_{n}\right)$ and performing the maximal cut, we find the top-sector reduction coefficient of $\bra{\varphi}$ onto $\bra{e_1}$ as
	\begin{equation}
		c_{1}=\frac{N(\bm{x}=\bm{0})}{D(\bm{x}=\bm{0})} \, .
	\end{equation}
	Note that here we don't need to compute any intersection numbers! We can now subtract the top-sector component at the integrand level:
	\begin{equation}
			\hat{\varphi}_{r} =\hat{\varphi}-c_1 \hat{e}_1 =\frac{N(\bm{x})D(\bm{x}=\bm{0}) - D(\bm{x})N(\bm{x}=0)}{x_1x_2...x_n \, D(\bm{x})D(\bm{x}=0)} \, .
	\end{equation}
	The numerator vanishes when $\bm{x} \to \bm{0}$, hence it contains no constant term. This means that the above subtracted integrand automatically have the sub-sector form. We can then perform the reduction recursively in the sub-sectors. 
	
	To summarize, for one-loop reduction we don't need to compute any intersection numbers at all, and we also don't need to perform the top-sector ISP reduction. The only operations are the transformations of the integrands to the regular form, which are as simple as taking a couple of derivatives. Everything else then follows directly.

	For illustration purpose, we show the example of the one-loop massless box family, which has also been used to demonstrate the top-down reduction in \cite{Frellesvig:2020qot}. The propagator denominators are given by
	\begin{equation}
		x_1=l_{1}^2 \,,\; x_2=(l_{1}-p_{1})^2 \,,\; x_3=(l_{1}-p_{1}-p_{2})^2 \,,\; x_4=(l_{1}-p_{1}-p_{2}-p_{3})^2 \, ,
	\end{equation}
	and the kinematic configuration is
	\begin{equation}
		p_{i}^2=0 \,,\; (i=1,2,3,4) \,, \quad (p_{1}+p_{2})^2=s \,, \quad (p_1+p_3)^2=t \, .
	\end{equation}
	The Baikov polynomial $P(\bm{x})$ is given by
	\begin{multline}
		P(\bm{x})=-s^2 x_2^2+2 t x_1 \left(x_3 (2 s+t)+s \left(t-x_4\right)-s x_2\right)-\left(s \left(x_4-t\right)+t x_3\right)^2 \\ +2 s x_2 \left(x_4 (s+2 t)+s t-t x_3\right)-t^2 x_1^2 \, .
	\end{multline}
	
	Now suppose that we want to reduce $I(3,2,1,1)$ in this family. We first transform the integrand of $I(3,2,1,1)$ to regular form:
	\begin{equation}\label{eq:3211}
		\hat{\varphi}=\frac{\mathcal{D}^{\vec{a}}P(\bm{x})^{\gamma}}{x_1x_2x_3x_4 \, P(\bm{x})^{\gamma}} \, ,
	\end{equation}
	where
	\begin{equation}
		\gamma=\frac{5-d}{2} \,, \quad \vec{a}=(3,2,1,1) \,, \quad  \mathcal{D}^{\vec{a}}=\frac{1}{2}\frac{\partial^2}{\partial x_1^2}\frac{\partial}{\partial x_2} \, .
	\end{equation}
	The master integrals can be chosen the same as in \cite{Frellesvig:2020qot}:
	\begin{equation}\label{eq:boxbasis}
		\hat{e}_1=\frac{1}{x_1x_2x_3x_4} \,, \quad \hat{e}_2=\frac{1}{x_1x_3} \,, \quad \hat{e}_3=\frac{1}{x_2x_4} \, ,
	\end{equation}
	where $\hat{e}_1$ is in the top sector and $\hat{e}_2,\hat{e}_3$ are sub-sector masters. We would like to know the coefficients $c_i$ in the decomposition
	\begin{equation}
		\lbra{\varphi}= c_1\lbra{e_1}+c_2\lbra{e_2}+c_3\lbra{e_3} \, .
	\end{equation}
	Performing the maximal cut, it is straightforward to obtain
	\begin{equation}\label{eq:c1}
		c_1=\frac{(d-7)(d-6)(d-5)}{2s^2t} \, .
	\end{equation}
	We then subtract the top-sector component to get $\hat{\varphi}_{r}=\hat{\varphi} - c_1/(x_1x_2x_3x_4)$. From the discussions before, we know that the integrand $\hat{\varphi}_r$ must automatically take the sub-sector form. Indeed, after taking into account the symmetries $x_1\leftrightarrow x_3$ and $x_2\leftrightarrow x_4$, the integrand can be recasted into the form
	\begin{equation}
		\hat{\varphi}_{r} \simeq \frac{N_{r1}(x_1,x_3)}{x_1x_2x_4 \, P(\bm{x})^3}+\frac{N_{r2}(x_2,x_4)}{x_1x_2x_3 \, P(\bm{x})^3} + \cdots \, ,
	\end{equation}
	where $N_{r1}(x_1,x_3)$ and $N_{r2}(x_2,x_4)$ are polynomials, and ``$\simeq$'' means equivalence after integration. The ellipsis denotes terms belonging to zero-sectors that vanish after integration. We discuss the identification of zero-sectors in Appendix~\ref{app:dimensionshift}. We will always drop these zero-sector terms in the following.
	
	We can now employ the recursion formula \eqref{eq:recursionformula} to integrate out $x_3$ for the first term and $x_4$ for the second term, respectively. The resulting expression automatically degenerate to sub-sectors $\{0,1,0,1\}$ and $\{1,0,1,0\}$. That is,
	\begin{equation}
		\hat{\varphi}_{r} \simeq \frac{N^{(1)}_{r1}(x_1)}{x_2x_4 \, P_{124}^{3}} + \frac{N^{(1)}_{r2}(x_2)}{x_1x_3 \, P_{123}^{3}} \, ,
	\end{equation}
	where $P_{124}$ is the Baikov polynomial in the representation for sector $\{1,1,0,1\}$ and $P_{123}$ is for $\{1,1,1,0\}$. We can then further integrate out $x_1$ and $x_2$ respectively for these two terms, and arrive at
	\begin{equation}
		\hat{\varphi}_{r} \simeq \frac{2t^2(2s+dt-8t)(d-7)(d-5)(d-3)}{(d-8)s^2 \, P_{24}^3}+\frac{2s^2(d-7)(d-5)(d-3)}{t \, P_{13}^3} \, ,
	\end{equation}
	where $P_{24}$ is the Baikov polynomial for sector $\{0,1,0,1\}$ and $P_{13}$ is for sector $\{1,0,1,0\}$. Their expressions are
	\begin{equation}
		\begin{aligned}
			P_{24}&=t^2+x_2^2+x_4^2-2tx_2-2tx_4-2x_2x_4 \, , \\
			P_{13}&=s^2+x_1^2+x_3^2-2sx_1-2sx_3-2x_1x_3 \, .
		\end{aligned}
	\end{equation}
	Performing maximal cut in these two sub-sectors, we find
	\begin{equation}
		c_2=\frac{2(d-7)(d-5)(d-3)}{s^4t} \,, \quad c_3=\frac{2(d-7)(d-5)(d-3)(2s+dt-8t)}{(d-8)s^2t^4} \, .
	\end{equation}
	Hence, we see that the complete reduction is achieved without computing any intersection numbers.

	\subsection{The unequal-mass sunrise family}\label{subsec:sunrise}

	\begin{figure}[t!]
		\centering
		\includegraphics[width=0.6\textwidth]{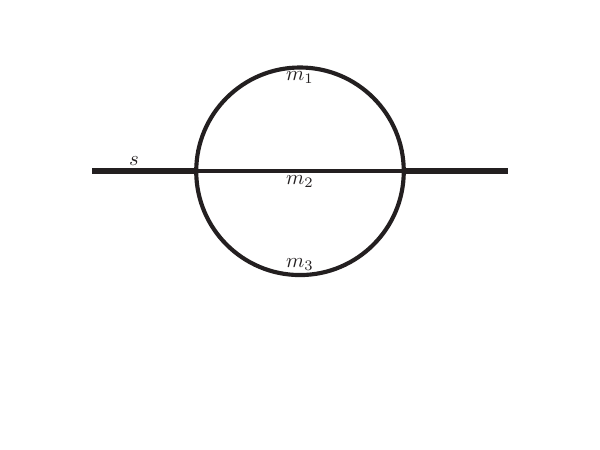}
		\vspace{-2.5cm}
		\caption{Unequal-mass sunrise family.}\label{fig:unequalmasssunrise}
	\end{figure}
		
	We now turn to a two-loop example which involves the top-sector ISP reduction. This is the unequal-mass sunrised family depicted in Fig.~\ref{fig:unequalmasssunrise}. The propagator denominators are given by
	\begin{equation}
		x_1=l_{1}^{2}-m_{1}^2 \,, \; x_2=(l_1-l_2)^2-m_{2}^{2} \,, \; x_3=(l_2-p)^2-m_{3}^{2} \,, \; x_4=l_{2}^2 \,, \; x_5=(l_{1}-p)^2 \, ,
	\end{equation}
	and the kinematic configuration is $p^2=s$. The $u$-function is $u(\bm{x})=P(\bm{x})^{-\epsilon}$ with the Baikov polynomial
	\begin{align}
		P(\bm{x}) &=s s_1s_2+ss_2s_3+ss_1s_3+s_1s_2s_3-s^2s_2-ss_2^2-s_1^2s_3-s_1s_3^2 +(s-s_1)(s_2-s_3)x_4 \nonumber
		\\
		&-(s_1-s_2)(s-s_3)x_5+(s+s_1+s_2+s_3)x_4x_5-(x_4+x_5)x_4x_5 \, ,
	\end{align}
	where $\bm{x}_{\prop} = \{x_1,x_2,x_3\}$ and $\bm{x}_{\isp}=\{x_4,x_5\}$. For later convenience, we define $s_i\equiv x_i+m_i^2$.
	
	To demonstrate our method, we consider the reduction of $I(1,1,1,-3,0)$ in this family. The integrand is already in the regular form:
	\begin{equation}
		\hat{\varphi} = \frac{x_4^3}{x_1x_2x_3} \, .
	\end{equation} 
	There are 4 master integrals in the top sector $\{1,1,1,0,0\}$. The number of master integrals is the same for both the maximal-cut reduction and the top-sector ISP reduction. We choose the following basis:
	\begin{equation}\label{eq:sunrisebasis1}
		\hat{e}_1=\frac{1}{x_1x_2x_3} \,,\; \hat{e}_2=\frac{x_4}{x_1x_2x_3} \,,\; \hat{e}_3=\frac{x_5}{x_1x_2x_3} \,,\; \hat{e}_4=\frac{x_4^2}{x_1x_2x_3} \, .
	\end{equation}
	We now perform the top-sector ISP reduction for $\hat{\varphi}$, which amounts to computing intersection numbers with $\bm{x}_{\prop}$ kept as constant. These 2-fold intersection numbers are straightforward to calculate, and we arrive at
	\begin{equation}\label{eq:fullreduction1}
		\lbra{\tilde{\varphi}}=\tilde{c}_1\lbra{\tilde{e}_1}+\tilde{c}_2\lbra{\tilde{e}_2}+\tilde{c}_3\lbra{\tilde{e}_3}+\tilde{c}_4\lbra{\tilde{e}_4}\, ,
	\end{equation}
	where
	\begin{align}
		\tilde{c}_1&=\frac{1}{3\epsilon-5}\left[s^2 \left(s_1 (\epsilon -2)-s_2 \epsilon \right)+s\left(s_1^2 (\epsilon -2)+s_2 \left(2 s_3-s_2 \epsilon\right)+2 \left(s_2+s_3\right) s_1\right) \right. \nonumber\\
		&\hspace{3em}\left.+s_3 \left(-s_1^2\epsilon +s_1 \left(2 s_2-s_3 \epsilon \right)+s_2\left(s_2+s_3\right) (\epsilon -2)\right)\right] \, ,\nonumber \\
		\tilde{c}_2&=\frac{1}{3\epsilon-5}\left[ -s^2(\epsilon -2)+s \left(s_1 (9-5 \epsilon )+s_2 (7-3\epsilon )+2 s_3 (\epsilon -2)\right)-s_1^2 (\epsilon-2) \right. \nonumber\\
		&\hspace{3em}\left.-s_2^2 \epsilon -s_3^2 \epsilon +2 s_1 s_2 (\epsilon-2)+s_1 s_3 (7-3 \epsilon )-5 s_2 s_3 \epsilon +2 s_2^2+2s_3^2+9 s_2 s_3\right] \, ,\nonumber \\
		\tilde{c}_3&=-\frac{2 \left(s_1-s_2\right) \left(s-s_3\right) (\epsilon -1)}{3\epsilon -5} \, ,\nonumber \\
		\tilde{c}_4&=\frac{\left(s+s_1+s_2+s_3\right) (4 \epsilon -7)}{3 \epsilon -5} \, .
	\end{align}
	The maximal cut $\bm{x}_{\prop} = 0$ corresponds to $s_1=m_1^2,s_2=m_2^2,s_3=m_3^2$. The limits of $\tilde{c}_i$ can be smoothly taken, and we recover the usual reduction coefficients as in Eq.~\eqref{eq:crelation}:
	\begin{equation}
		c_{i} = \tilde{c}_{i}(\bm{x}_{\prop}=0) \,, \; (i=1,2,3,4) \,.
	\end{equation}
	
	We can then subtract the top-sector components of $\lbra{\varphi}$ and use \eqref{eq:remainingintsub} to reduce the integrand to sub-sectors. We have 
	\begin{equation}\label{eq:uneqmasssunrise_phir}
			\hat{\varphi}_{r}=\hat{\varphi} - \sum_{i=1}^{4}c_i \hat{e}_i 
			\simeq \frac{N_{r1}(x_3,x_4,x_5)}{x_1x_2}+\frac{N_{r2}(x_1,x_4,x_5)}{x_2x_3}+\frac{N_{r3}(x_2,x_4,x_5)}{x_1x_3} \, .
	\end{equation}
	The last two terms can be obtained from the first one by the substitutions:
	\begin{equation}\label{eq:symuneqmasssunrise}
			m_1^2 \leftrightarrow m_3^2 \,,\; m_2^2 \leftrightarrow s \,,\; x_1 \rightarrow x_3 \, ; \quad
			m_1^2 \leftrightarrow s \,,\; m_2^2 \leftrightarrow m_3^2 \,,\; x_2 \rightarrow x_3 \, .
	\end{equation}
	Therefore it is enough to consider the first term. The explicit expression for the numerator can be written as
	\begin{equation}
		\begin{aligned}
			&N_{r1}(x_3,x_4,x_5) = \frac{1}{3\epsilon-5}\left[ 2 m_1^2 s+2 m_2^2 s-m_1^4 \epsilon -2 m_3^2 m_1^2 \epsilon +m_2^4 \epsilon +2 m_2^2 m_3^2 \epsilon \right. \\
			&+2 m_2^2 m_1^2-2 m_2^4-4 m_2^2 m_3^2 -\left(3 m_1^2 \epsilon +5 m_2^2 \epsilon +2 m_3^2 \epsilon -7 m_1^2-9 m_2^2-4 m_3^2-2 s \epsilon +4 s\right)x_4 \\
			&\left.+\left(-2m_2^2-m_1^2\epsilon+m_2^2\epsilon\right)x_3 +2(m_1^2-m_2^2)(\epsilon-1)x_5-(\epsilon-2)x_3x_4+(4\epsilon-7)x_4^2 \right] \, .
		\end{aligned}
	\end{equation}
	By integrating out $x_3$, $x_4$ and $x_5$, the first term in Eq.~\eqref{eq:uneqmasssunrise_phir} can be directly reduced to the sub-sector $\{1,1,0,0,0\}$. There is only one master integral in this sub-sector, which we choose as $e_5 = I(1,1,0,0,0)$. The corresponding coefficient $c_5$ can be easily obtained from maximal cut, and is given by
	\begin{equation}
		c_5 = -\frac{2 m_1^2 (\epsilon -1) \left(2 m_2^2 \epsilon +m_3^2 \epsilon -3 m_2^2-2 m_3^2-s\epsilon +2 s\right)}{(\epsilon -2) (3 \epsilon -5)} \, .
	\end{equation}
	Using the substitutions in \eqref{eq:symuneqmasssunrise}, the other two terms in \eqref{eq:uneqmasssunrise_phir} can be reduced to the master integrals $e_6 = I(0,1,1,0,0)$ and $e_7 = I(1,0,1,0,0)$. The corresponding coefficients are
	\begin{equation}
		\begin{aligned}
			c_6 =& -\frac{2 m_3^2 (\epsilon -1) \left(m_1^2 \epsilon -m_2^2 \epsilon -2 m_1^2+2 m_2^2+2 s \epsilon -3 s\right)}{(\epsilon -2) (3 \epsilon -5)} \, , \\
			c_7 =& -\frac{2 s (\epsilon -1) \left( -m_1^2 \epsilon +m_2^2 \epsilon +2 m_3^2 \epsilon +2m_1^2-2 m_2^2-3 m_3^2\right)}{(\epsilon -2) (3 \epsilon -5)}\, .
		\end{aligned}
	\end{equation}
	This completes the reduction. The reduction coefficients $c_i, (i=1,\ldots,7)$ can be compared to the results from \texttt{Kira}, and we find complete agreement.
	
	Integrals with higher powers of propagators can be reduced similarly. We use $I(1,1,3,0,0)$ as an example. Furthermore, we choose the master integrals to be
	\begin{equation}\label{eq:sunrisebasis2}
		\hat{e}_1 = \frac{1}{x_1^2x_2x_3} \,, \; \hat{e}_2=\frac{1}{x_1x_2^2x_3} \,, \; \hat{e}_3=\frac{1}{x_1x_2x_3^2} \,, \; \hat{e}_4=\frac{1}{x_1x_2x_3} \, ,
	\end{equation}
	which also exhibit higher powers in the denominators. To proceed, we first transform all the integrands to regular form defined in \eqref{eq:regularform}. They become
	\begin{equation}
		\begin{aligned}
			\hat{\varphi}&=\frac{1}{2x_1x_2x_3 \, P(\bm{x})^{-\epsilon}}\frac{\partial^2 P(\bm{x})^{-\epsilon}}{\partial^2 x_3} \, , \\
			\hat{e}_1&=\frac{1}{x_1x_2x_3 \, P(\bm{x})^{-\epsilon}}\frac{\partial P(\bm{x})^{-\epsilon}}{\partial x_1} \, , \quad \hat{e}_2=\frac{1}{x_1x_2x_3 \, P(\bm{x})^{-\epsilon}}\frac{\partial P(\bm{x})^{-\epsilon}}{\partial x_2} \, , \\
			\hat{e}_3&=\frac{1}{x_1x_2x_3 \, P(\bm{x})^{-\epsilon}}\frac{\partial P(\bm{x})^{-\epsilon}}{\partial x_3} \, , \quad \hat{e}_4=\frac{1}{x_1x_2x_3} \, .
		\end{aligned}
	\end{equation}
	Performing the top-sector ISP reduction, we get
	\begin{equation}\label{eq:fullreduction2}
		\lbra{\tilde{\varphi}}=\tilde{c}_1\lbra{\tilde{e}_1}+\tilde{c}_2\lbra{\tilde{e}_2}+\tilde{c}_3\lbra{\tilde{e}_3}+\tilde{c}_4\lbra{\tilde{e}_4}\, , \quad \tilde{c}_{i}=\frac{N_{1}^{(i)}(\bm{x}_{\prop})}{D_{1}(\bm{x}_{\prop})} \, .
	\end{equation}
	The explicit expression of $D_{1}$ is
	\begin{align}
		D_{1}=&s_3 \left(s^4-4 \left(s_1+s_2+s_3\right) s^3+\left(6 s_1^2+4\left(s_2+s_3\right) s_1+6 s_2^2+6 s_3^2+4 s_2 s_3\right) s^2 \right. \nonumber \\
		&-4\left(s_1^3-\left(s_2+s_3\right) s_1^2-\left(s_2^2-10 s_3 s_2+s_3^2\right)s_1+\left(s_2-s_3\right){}^2 \left(s_2+s_3\right)\right) s \nonumber \\
		&\left.+\left(s_1^2-2\left(s_2+s_3\right) s_1+\left(s_2-s_3\right){}^2\right){}^2\right) \, ,
	\end{align}
	and $N_1^{(i)}$ can be found in Appendix~\ref{app:sunrise}.
	Taking the maximal cut, i.e., setting $s_i$ to $m_{i}^2$, we have
	\begin{equation}
		D_{0}=D_{1}(\bm{x}_{\prop}=\bm{0}) \,, \quad N^{(i)}_{0}=N^{(i)}_{1}(\bm{x}_{\prop}=\bm{0}) \, .
	\end{equation}
	The ratios $c_i=N^{(i)}_{0}/D_{0}$ are essentially the reduction coefficients in the usual IBP reduction. Applying \eqref{eq:generalreduction}, we find that the top-sector subtracted integrand can be transformed into three sub-sectors:
	\begin{equation}
		\label{eq:sunrisebasis3}
		\varphi_{r}\simeq \frac{N_{r1}(x_1,x_2,x_3)}{D_{0}x_1^2x_2^2}+\frac{N_{r2}(x_1,x_2,x_3)}{D_{0}x_1^2x_3^3}+\frac{N_{r3}(x_1,x_2,x_3)}{D_{0}x_2^2x_3^3} \, .
	\end{equation}
	After integrating out the ISPs in each sub-sector, we can perform the reductions under maximal cuts. The master integrals for these sub-sectors can be chosen as $I(1,1,0,0,0)$, $I(0,1,1,0,0)$ and $I(1,0,1,0,0)$, respectively. The reduction coefficients are
	\begin{equation}
		\begin{aligned}
			c_5 =& \frac{(1\!-\!\epsilon)^2\left(m_1^4\!-\!2 m_1^2 s\!-\!2 m_2^2 s+2 m_3^2 s-2 m_2^2 m_1^2+2 m_3^2 m_1^2+m_2^4-3m_3^4+2 m_2^2 m_3^2+s^2\right)}{D_{0}} \, , \\
			c_6 =& \frac{(1\!-\!\epsilon)^2\left(2 m_1^2 s\!-\!2 m_2^2 s\!-\!2 m_3^2 s-3 m_1^4+2 m_2^2 m_1^2+2 m_3^2m_1^2+m_2^4+m_3^4-2 m_2^2 m_3^2+s^2\right)}{D_{0}} \, , \\
			c_7 =& \frac{(1\!-\!\epsilon)^2\left(m_1^4\!-\!2 m_1^2 s\!+\!2 m_2^2 s-2 m_3^2 s+2 m_2^2 m_1^2-2 m_3^2 m_1^2-3m_2^4+m_3^4+2 m_2^2 m_3^2+s^2\right)}{D_{0}} \, ,
		\end{aligned}
	\end{equation}
	where
	\begin{equation}
		\begin{aligned}
			D_{0}=&m_3^2 \left(m_1^4 \left(4 m_2^2 \left(m_3^2+s\right)+4 m_3^2 s+6 m_2^4+6m_3^4+6 s^2\right) \right. \\
			&-4 m_1^2 \left(-m_2^2 \left(-10 m_3^2s+m_3^4+s^2\right)-m_2^4 \left(m_3^2+s\right)+\left(m_3^2-s\right){}^2\left(m_3^2+s\right)+m_2^6\right)  \\
			&\left.-4 m_1^6\left(m_2^2+m_3^2+s\right)+\left(-2 m_2^2\left(m_3^2+s\right)+\left(m_3^2-s\right){}^2+m_2^4\right){}^2+m_1^8\right) \, .
		\end{aligned}
	\end{equation}
	Again, the reduction coefficients agree perfectly with the usual IBP reduction from \texttt{Kira}.

\subsection{The equal-mass sunrise family}

	It is interesting to study the equal-mass case of the above sunrise family, where $m_1^2=m_2^2=m_3^2=m^2$. This case is simpler with degenerate kinematics, and it may appear that one can easily obtain the results by taking the limit from the more general unequal-mass case. However, we will show that one needs to be careful regarding the increased symmetry of the degenerate case.

	In the previous subsection, we have employed two kinds of bases for the unequal-mass family, Eqs.~\eqref{eq:sunrisebasis1} and \eqref{eq:sunrisebasis2}. In the equal-mass limit, there is a symmetry with respect to the exchange among the three propagators. One can see that the basis in Eq.~\eqref{eq:sunrisebasis2} explicitly encodes this symmetry, such that $\rbra{e_1}=\rbra{e_2}=\rbra{e_3}$ in the degenerate limit. The same is true for the sub-sector master integrals introduced below Eq.~\eqref{eq:sunrisebasis3}. Therefore, it is straightforward to take the limit and obtain the reduction result for the equal-mass sunrise family:
	\begin{equation}
		\rbra{\varphi}=(c_1+c_2+c_3)\rbra{e_1}+c_4\rbra{e_4}+(c_5+c_6+c_7)\rbra{e_5} \, .
	\end{equation}

	On the other hand, if we take the basis of Eq.~\eqref{eq:sunrisebasis1}, the degenerate symmetry is somewhat hidden. In fact, we can see that $\rbra{e_2}=\rbra{e_3}$ in the equal-mass limit, and obtain
	\begin{equation}
		\rbra{\varphi}=c_1\rbra{e_1}+(c_2+c_3)\rbra{e_2}+c_4\rbra{e_4}+(c_5+c_6+c_7)\rbra{e_5} \, .
	\end{equation}	
	However, in this limit, the integral $\rbra{e_4}$ is also related to the other master integrals and is hence reducible. The reducibility of $\rbra{e_4}$ has a connection to the increased symmetry, but the reduction coefficients can not be easily seen. Therefore, the lesson to be learnt here is that one needs to choose the basis carefully to maximally exploit the symmetries of the integral family.

	\subsection{The 3-loop banana integral family}\label{subsec:banana}

	\begin{figure}[t]
		\centering
		\includegraphics[width=0.6\textwidth]{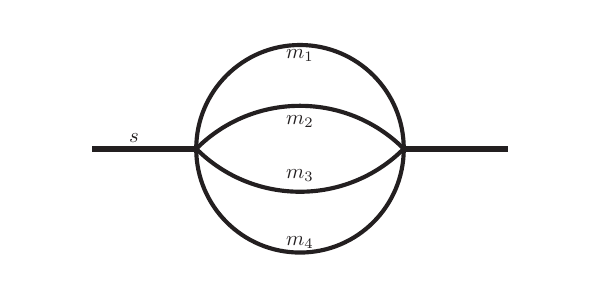}
		\caption{Three-loop unequal-mass banana integral family.}\label{fig:banana}
	\end{figure}
	
	The method extends to higher loop orders as well. We study in this section the 3-loop banana integral family. The topology is depicted in Fig.~\ref{fig:banana}, with the propagator denominators given by
	\begin{equation}
		\begin{aligned}
			x_1&=k_1^2-m_1^2 \,,\; x_2=k_2^2-m_2^2 \,,\; x_3=(k_1-k_3)^2-m_3^2 \,,\; x_4=(k_2-k_3-p)^2-m_4^2 \,, \\
			x_5&=k_3^2 \,,\; x_6=(k_1-p)^2 \,,\; x_7=(k_2-p)^2 \,,\; x_8=(k_3-p)^2 \,,\; x_9=(k_1-k_2)^2 \, ,
		\end{aligned}
	\end{equation}
	where $p^2=s$, $\bm{x}_{\prop}=\{x_1,x_2,x_3,x_4\}$, $\bm{x}_{\isp}=\{x_5,x_6,x_7,x_8,x_9\}$.
	
	We start from the standard representation with $u_{\text{std}}=G(k_1,k_2,k_3,p)$, whose explicit expression is too cumbersome to be shown in the paper. We sequentially integrate out $x_9$, $x_6$ and $x_8$ to arrive at a loop-by-loop representation for the top sector, where the $u$-function is given by (up to an irrelevant constant factor)
	\begin{equation}\label{eq:bananatoprep}
		u_{\text{LBL}} = x_5^{\epsilon} \, x_7^{\epsilon} \, \lambda(x_5,s_1,s_3)^{-1/2-\epsilon} \, \lambda(x_5,x_7,s_4)^{-1/2-\epsilon} \, \lambda(x_7,s,s_2)^{-1/2-\epsilon} \, ,
	\end{equation}
	where $s_{i}\equiv x_i+m_i^{2}$, and $\lambda(x,y,z)$ is the K\"all\'en function defined as 
	\begin{equation}
		\lambda(x,y,z)\equiv x^2+y^2+z^2-2xy-2yz-2zx \, .
	\end{equation}
	
	We consider the integral $I(1,1,1,1,0,-1,0,0,0)$, whose integrand in the loop-by-loop representation \eqref{eq:bananatoprep} can be obtained by integrating out $x_9$, $x_6$ and $x_8$, starting from the standard one $x_6/(x_1 x_2 x_3 x_4)$. The result is given by
	\begin{equation}
	    \label{eq:integrand_banana}
		\begin{aligned}
			\hat{\varphi}&= \frac{1}{x_1 x_2 x_3 x_4} \, \frac{N_{\varphi}}{4x_5x_7} \, , \\
			N_{\varphi}&= s x_5^2-s_2 x_5^2+s s_1 x_5-s_1 s_2 x_5-s s_3 x_5+s_2 s_3 x_5-s s_4 x_5+s_2 s_4 x_5 +5 s x_7 x_5 \\ 
			&+5 s_1 x_7 x_5-s_2 x_7 x_5-s_3 x_7 x_5-s_4 x_7 x_5+s_1 x_7^2 -s_3 x_7^2+s s_1 x_7 \\
			&-s_1 s_2 x_7-s s_3 x_7+s_2 s_3 x_7-s_1 s_4 x_7+s_3 s_4 x_7 -s s_1 s_4+s_1 s_2 s_4+s s_3 s_4 \\
			&-s_2 s_3 s_4+x_7x_5^2+x_7^2 x_5 \, .
		\end{aligned}
	\end{equation}
	Note that, comparing to Eq.~\eqref{eq:phi_with_dots}, the integrand here contains extra factors (i.e., $x_5$ and $x_7$) in the denominator. This is common when working with non-standard Baikov representations at multi-loop levels. These factors are present in the $u$-function above, and hence correspond to twisted boundaries. To reconcile this kind of integrands with the derivations following Eq.~\eqref{eq:phi_with_dots}, it is enough to extend the function $Q$ in \eqref{eq:phi_with_dots} to rational functions of $\bm{x}$, whose singularities only live on twisted boundaries.
	
	There are 11 master integrals in the top sector.\footnote{This number can be correctly computed from the dimension of the cohomology group in the standard representation. In the loop-by-loop representation \eqref{eq:bananatoprep}, the dimension will be higher, since the representation allows extra integrals with $x_5$ and/or $x_7$ in the denominator.} The number is the same for top-sector ISP reduction. We choose the master integrals to be
	\begin{align}
		&e_{1}:\, I(1,1,1,1,0,0,0,0,0) \,,\; e_{2}: \, I(2,1,1,1,0,0,0,0,0) \,,\; e_{3}: \,I(1,2,1,1,0,0,0,0,0) \,, \nonumber \\ 
		&e_{4}:\, I(1,1,2,1,0,0,0,0,0) \,,\; e_{5}:\,I(1,1,1,2,0,0,0,0,0) \,,\; e_{6}:\,I(2,1,1,2,0,0,0,0,0) \,, \nonumber \\
		&e_{7}:\,I(2,1,2,1,0,0,0,0,0) \,,\; e_{8}:\,I(1,2,1,2,0,0,0,0,0) \,,\; e_{9}:\,I(1,2,2,1,0,0,0,0,0) \,,  \nonumber \\
		&e_{10}:\,I(1,1,2,2,0,0,0,0,0) \,,\; e_{11}:\,I(1,1,1,3,0,0,0,0,0) \,.
	\end{align}
	Performing the top-sector ISP reduction, we get
	\begin{equation}\label{eq:bananatop}
		\lbra{\tilde{\varphi}}= \sum_{i=1}^{11}\tilde{c}_{i}\lbra{\tilde{e}_i} \, ,
	\end{equation}
	where $\tilde{c}_{i}$ are functions of $\epsilon$, $s$ and $s_{i}$, whose explicit expressions are given in Appendix.~\ref{app:banana}. Taking the limit $s_{i}\rightarrow m_{i}^{2}$, we get the maximal-cut reduction coefficients $c_i$:
	\begin{equation}
		\tilde{c}_{i}\equiv \frac{N^{(i)}_{1}}{D_{1}} \,,\quad \tilde{c}_{i}\Big|_{s_{i}\rightarrow m_{i}^2}=c_{i}\equiv \frac{N^{(i)}_{0}}{D_{0}} \, .
	\end{equation}
	
	We can now construct the top-sector subtracted integrand according to Eq.~\eqref{eq:generalreduction}. At this point, we note that one prominent feature of the above reduction coefficients is
	\begin{equation}\label{eq:bananaD1}
		D_{1}=(-1+\epsilon)^2 = D_{0} \, ,
	\end{equation}
	which means that $I^{(0)}_{r}$ in \eqref{eq:generalreduction} vanishes. This is important since, according to the discussions below Eq.~\eqref{eq:integrand_banana}, there are extra factors of $1/x_5$ and $1/x_7$ in the $Q$-function inside the definition of $I^{(0)}_{r}$. If $I^{(0)}_{r}$ is not 0, these $1/x_5$ and $1/x_7$ factors will remain in the subtracted integrand $\hat{\varphi}_r$. We will encounter this kind of situations in Sec.~\ref{sec:reductionm}, and discuss methods to deal with these more complicated cases. In the current case, the subtracted integrand can be transformed into 4 sub-sectors:
	\begin{equation}
		\varphi_{r}\simeq \frac{N_{r1}(x_2,x_3)}{x_1x_2^2x_3^2}+\frac{N_{r2}(x_2,x_3,x_4)}{x_1x_2^2x_4^3}+\frac{N_{r3}(x_2,x_3,x_4)}{x_1x_3^2x_4^3}+\frac{N_{r4}(x_1,x_2,x_3,x_4)}{x_2^2x_3^2x_4^3} \, .
	\end{equation}
	These 4 terms can be dealt with in 4 different representations. They can be obtained sequentially integrating out variables as following:
	\begin{align*}
	u_{\text{LBL}} &\xrightarrow{\text{integrating out $x_4,x_5,x_7$}} u_{r1}(x_1,x_2,x_3) \,,
	\\
	u_{\text{LBL}} &\xrightarrow{\text{integrating out $x_3,x_5,x_7$}} u_{r2}(x_1,x_2,x_4) \,,
	\\
	u_{\text{LBL}} &\xrightarrow{\text{integrating out $x_2,x_7,x_5$}} u_{r3}(x_1,x_3,x_4) \,,
	\\
	u_{\text{LBL}} &\xrightarrow{\text{integrating out $x_1,x_5,x_7$}} u_{r4}(x_2,x_3,x_4) \,.
	\end{align*}
	The four sub-sector master integrals can be chosen as
	\begin{equation}
		\begin{aligned}
			&e_{12} = I(1,1,1,0,0,0,0,0,0) \,, \quad e_{13} = I(1,1,0,1,0,0,0,0,0) \,,\\ 
			&e_{14} = I(1,0,1,1,0,0,0,0,0) \,, \quad e_{15} = I(0,1,1,1,0,0,0,0,0) .
		\end{aligned}
	\end{equation}
	The reduction coefficients can be easily obtained from maximal cuts, and are given by
	\begin{equation}
	c_{12} = -\frac{1}{6} \,, \quad c_{13} = -\frac{1}{6} \,, \quad c_{14} = -\frac{1}{6} \,, \quad c_{15} = \frac{3}{2} \,.
	\end{equation}

	We see that the top-sector ISP reduction for the banana family is more complicated than the sunrise family. However, after subtracting the top-sector components, the reductions in the sub-sectors are almost trivial for both families. This is of course a special feature of the banana and sunrise families, since the sub-sectors are all products of one-loop tadpoles. This simplicity should not be expected in general multi-loop families, as we will see in the next Section.

	\section{Top-sector ISP reduction for more general cases}\label{sec:reductionm}
	
	In this Section, we discuss the top-sector ISP reduction for more general multi-loop families. We use the massless and massive double-box families as examples. We demonstrate that our method still works when the number of master integrals in top-sector ISP reduction is larger than the maximal-cut case. We also discuss some complications when performing the recursive reduction in sub-sectors.
		
	\subsection{The massless double-box family}

	\begin{figure}[t]
		\centering
		\includegraphics[width=0.6\textwidth]{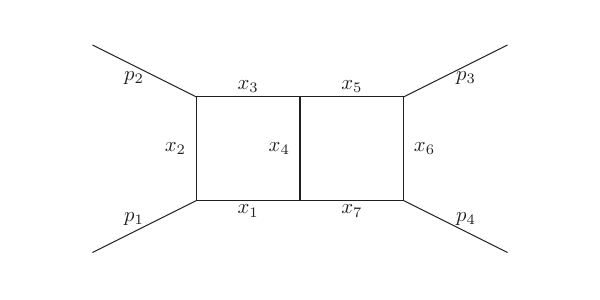}
		\caption{Massless double box family.}\label{fig:masslessdb}
	\end{figure}
	
	This family is depicted in Fig.~\ref{fig:masslessdb}, and the propagator denominators are given by
	\begin{equation}
	    \label{eq:propagators_masslessdb}
		\begin{aligned}
			x_1&=l_1^2 \,,\; x_2=(l_1-p_1)^2 \,,\; x_3=(l_1-p_1-p_2)^2 \,,\; x_4=(l_1-l_2)^2 \,,\; x_5=(l_2-p_1-p_2)^2 \,, \\
			x_6&=(l_2-p_1-p_2-p_3)^2 \,,\; x_7=l_2^2 \,,\; x_8=(l_2-p_1)^2 \,,\; x_9=(l_1-p_1-p_2-p_3)^2 \,,
		\end{aligned}
	\end{equation}
	where the kinematic configuration is
	\begin{equation}
		p_i^2=0 \,, (i=1,2,3,4) \,, \quad (p_1+p_2)^2=s \,, \quad (p_2+p_3)^2=t \, .
	\end{equation}
	Starting from the standard representation, we integrate out $x_9$ to arrive a loop-by-loop representation. The $u$-function is
	\begin{equation}\label{eq:dbtopbaikov}
		\begin{aligned}
			u&=P_{1}^{\epsilon} \, P_{2}^{-1/2-\epsilon} \, P_{3}^{-1/2-\epsilon} \, , \\
			P_{1}&=-4G(l_2,p_1,p_2)/s \,, \quad  P_{2}=16G(l_2,p_1,p_2,p_3) \,, \quad  P_{3}=16G(l_1,l_2,p_1,p_2) \, . 
		\end{aligned}
	\end{equation}
	The 3 polynomials are complicated functions in terms of the variables $x_i$. To simplify the expressions, we introduce a set of new variables (hinted by the form of the Gram matrices):
	\begin{equation}\label{eq:zvariables}
		\begin{aligned}
			z_1&\equiv x_1 \,,\; z_2\equiv x_1-x_2 \,,\; z_3\equiv s+x_2-x_3 \,,\; z_4\equiv x_1-x_4+x_7 \,,\; z_5\equiv s-x_5+x_7 \,, \\
			z_6&\equiv -s+x_5-x_6 \,,\; z_7\equiv x_7 \,,\; z_8\equiv s-x_5+x_8 \,. 
		\end{aligned}
	\end{equation}
	The top-sector ISP reduction involves intersection numbers over the ISP $x_8$, while keeping $x_1,\ldots,x_7$ fixed. This can be turned into intersection numbers over the new variable $z_8$, while keeping $z_1,\ldots,z_7$ fixed (since they do not depend on $x_8$). This kind of variable changes can often greatly simplify the expressions and accelerate the calculations. To simplify further, we set $s=1$ and recover it in the end by dimension counting. The explicit expressions of the polynomials are then given by
	\begin{equation}
		\begin{aligned}
			P_{1}&=z_7-z_5z_8+z_8^2 \, , \\
			P_{2}&=t^2 z_5^2-4 t^2 z_7-2 t z_5 z_6-4 t z_7+z_6^2+2 z_8 \left(t z_5+2 t z_6+z_6\right)+z_8^2 \, , \\
			P_{3}&=z_4^2-2 z_3 z_5 z_4+z_3^2 z_5^2-4 z_1 z_7+4 z_2 z_3 z_7+\left(z_2^2+2 z_3 z_2+z_3^2-4 z_1\right) z_8^2 \\
			&-2 \left(z_5 z_3^2-z_4 z_3+z_2 z_5 z_3+z_2 z_4-2 z_1 z_5\right) z_8 \, .
		\end{aligned}
	\end{equation}
	
	We now consider the reduction of the integral
	\begin{equation}
	    \varphi = I(1,1,1,1,1,1,1,-2,0) \,, \quad \hat{\varphi}=\frac{x_8^2}{x_1x_2x_3x_4x_5x_6x_7} \,,
	\end{equation}
	in this family. The number of master integrals in the top sector is 2, which can be obtained by computing the dimension under maximal cut. However, the dimension of the cohomology group is 5 in the top-sector ISP reduction. This is a new phenomenon that did not happen in the sunrise and banana families. We will show that our method still works in this situation, as mentioned below Eq.~\eqref{eq:fullreduction}.
	
	We choose the following ISP-integrated integrals as the basis for top-sector ISP reduction:
	\begin{align}\label{eq:doublebox_firstbasis}
			\tilde{e}_1 &= \tilde{I}(1,1,1,1,1,1,1,0,0) \,,\; \tilde{e}_2 = \tilde{I}(1,1,1,1,1,1,1,-1,0) \,,\; \tilde{e}_3 = \tilde{I}(2,1,1,1,1,1,1,0,0) \,, \nonumber \\ 
			\tilde{e}_4 &= \tilde{I}(1,1,2,1,1,1,1,0,0) \,,\; \tilde{e}_5 = \tilde{I}(1,1,1,1,2,1,1,0,0) \,.
	\end{align}
	It is clear that after fully-integrating over the remaining variables $\bm{x}_{\prop}$, three of the above basis become reducible. For the choice of basis, we have employed the symmetries under exchanges of variables, such that after full-integration, $\bra{e_3} = \bra{e_4} = \bra{e_5}$. They can then be reduced as
	\begin{equation}
		\lbra{e_3}=\frac{1+2\epsilon}{s}\lbra{e_1}+ \text{sub-sector integrals} \, ,
	\end{equation}
	which can be easily obtained under maximal cut.
	
	We transform the integrands in \eqref{eq:doublebox_firstbasis} to the regular form:
	\begin{align}\label{eq:doublebox_firstbasis_integrand}
			\hat{e}_1&=\frac{1}{x_1x_2x_3x_4x_5x_6x_7} \,,\; \hat{e}_2=\frac{x_8}{x_1x_2x_3x_4x_5x_6x_7} \,,\; \hat{e}_3=\left(-\frac{1}{2}-\epsilon\right)\frac{\partial_{x_1}P_{3}}{P_{3}}\frac{1}{x_1x_2x_3x_4x_5x_6x_7} \, , \nonumber \\
			\hat{e}_4&=\left(-\frac{1}{2}-\epsilon\right)\frac{\partial_{x_3}P_{3}}{P_{3}}\frac{1}{x_1x_2x_3x_4x_5x_6x_7}\, , \nonumber \\
			\hat{e}_5&=\left[\epsilon\frac{\partial_{x_5}P_1}{P_{1}}+\left(-\frac{1}{2}-\epsilon\right)\frac{\partial_{x_5}P_{2}}{P_{2}}+\left(-\frac{1}{2}-\epsilon\right)\frac{\partial_{x_5}P_{3}}{P_{3}}\right]\frac{1}{x_1x_2x_3x_4x_5x_6x_7} \, .
	\end{align}
	We then perform the top-sector ISP reduction for $\bra{\tilde{\varphi}}$ and obtain
	\begin{equation}\label{eq:doubleboxfullred}
		\lbra{\tilde{\varphi}}=\tilde{c}_1\lbra{\tilde{e}_1}+\tilde{c}_2\lbra{\tilde{e}_2}+\tilde{c}_3\lbra{\tilde{e}_3}+\tilde{c}_4\lbra{\tilde{e}_4}+\tilde{c}_5\lbra{\tilde{e}_5} \, ,
	\end{equation}
	where $\tilde{c}_{i}=N_{1}^{(i)}/D_{1}$ and
	\begin{equation}
		\begin{aligned}
			D_{1}&=(-1+2\epsilon)Q_1Q_2Q_3Q_4 \, . \\
			Q_{1}&=(s-x_5+2x_6-x_7) \, , \\
			Q_{2}&=\left(s^2-2 s x_1-2 s x_3+x_1^2+x_3^2-2 x_1 x_3\right) \, , \\
			Q_{3}&=\left(s x_7+s x_6-x_7^2+x_5 x_7+x_6 x_7-x_5 x_6\right) \, , \\
			Q_{4}&=\left(x_1 x_2-x_3 x_2+x_5 x_2-x_7 x_2-x_1 x_4+x_3 x_4+x_1 x_5-x_3 x_7\right) \, .
		\end{aligned}
	\end{equation}
	The numerators $N_{1}^{(i)}$ are complicated and we don't list their explicit expressions here. We now need to take the maximal-cut limit $\bm{x}_{\prop} \to \bm{0}$. It turns out that only $\tilde{c}_2$ has a well-defined multivariate limit:
	\begin{equation}
		c_2 = \lim_{\bm{x}_{\prop} \to \bm{0}} \tilde{c}_{2}=\frac{t+3s\epsilon}{1-2\epsilon} \, .
	\end{equation}
	For the remaining coefficients, we note that $\bra{e_3}$, $\bra{e_4}$ and $\bra{e_5}$ are reducible to $\bra{e_1}$ under maximal cut. Therefore, we only need to consider the limit of the combination
	\begin{equation}
	    c_1 = \lim_{\bm{x}_{\prop} \to \bm{0}} \left( \tilde{c}_1 + \frac{1+2\epsilon}{s} ( \tilde{c}_3 + \tilde{c}_4 + \tilde{c}_5 ) \right) = - \frac{s t \epsilon}{1-2\epsilon} \,.
	\end{equation}
	We have checked that $c_1$ and $c_2$ correctly reproduce the reduction coefficients of $\bra{\varphi}$ onto $\bra{e_1}$ and $\bra{e_2}$.

	We now proceed to perform the top-sector subtraction, and transform the subtracted integrand to sub-sectors. For this we would like to apply Eq.~\eqref{eq:generalreduction}. However, since $D_1(\bm{x}_{\prop} \to \bm{0}) = 0$, the $C_0$ in \eqref{eq:generalreduction} and \eqref{eq:generalreduction_C0} is not well-defined. To get around this, we introduce a new function $D'_0(\bm{x}_{\prop})$ to replace $D_0$ in \eqref{eq:generalreduction}, and define $C_0$ as the maximal-cut limit of $D'_0/D_1$. Ideally speaking, we would like $D'_0$ to satisfy two conditions: 1) it should be a monomial of the Baikov variables, as it appears in the denominator of \eqref{eq:generalreduction}, and we do not want to introduce extra polynomial factors there; 2) the limit $\bm{x}_{\prop} \to \bm{0}$ of $D'_0/D_1$ exists. However, in general these two conditions cannot be satisfied simultaneously. So we weaken the second condition, and only require the existence of the directional limit along a particular path. 
	
	There is some freedom in the choice of the path. The final result is independent of the path, as long as the same one is taken everywhere. To be concrete, we choose to take $x_1,x_3,x_4,x_5,x_6$ to 0 first, and then take $x_2$ and $x_7$ to 0. We can see that
	\begin{equation}
		D_{1} \xrightarrow{(x_1,x_3,x_4,x_5,x_6) \to 0} (1-2\epsilon)s^2(s-x_7)^2x_2x_7^2 \, .
	\end{equation}
	Therefore, we can choose
	\begin{equation}
			D_0^{\prime}=(1-2\epsilon)s^4x_2x_7^2 \,,
	\end{equation}
	and define
	\begin{align}
	    C_0 &= \lim_{x_2,x_7 \to 0} \frac{D_0^{\prime}}{D_{1}((x_1,x_3,x_4,x_5,x_6) \to 0)} = 1 \,, \nonumber
	    \\
	     {N_{0}^{(i)}}^{\prime}&= \frac{N_{0}^{(i)}}{D_0} \, D_0^\prime = N_{0}^{(i)} \, s^4x_2x_7^2 \,,
	\end{align}
	where $D_0=1-2\epsilon$. We can now replace $D_{0},N^{(i)}_{0}$ in Eq.~\eqref{eq:generalreduction} by $D_{0}^{\prime}$ and ${N_{0}^{(i)}}^{\prime}$, and obtain
	\begin{align}
		I_{r}&=I^{(0)}_{r}+\sum_{i}I^{(i)}_{r} \nonumber \\
		I_{r}^{(0)}&=(-1)^{|a|-n} \int_{\mathcal{C}} \mathrm{d}^n \bm{x}_{\prop} \, \mathrm{d}^{m} \bm{x}_{\isp} \, u(\bm{x}) \, Q(\bm{x}_{\isp}) \, \mathcal{D}^{\vec{a}}\frac{D_{0}^\prime-D_{1}(\bm{x}_{\prop})}{D_{0}^\prime \, x_1 \cdots x_n} \,, \nonumber
		\\
		I_{r}^{(i)}&=(-1)^{|b^{(i)}|-n+1} \int_{\mathcal{C}} \mathrm{d}^n \bm{x}_{\prop} \, \mathrm{d}^{m} \bm{x}_{\isp} \, u(\bm{x}) \, Q_i(\bm{x}_{\isp}) \, \mathcal{D}^{\vec{b}^{(i)}} \frac{{N^{(i)}_{0}}^{\prime}-N^{(i)}_{1}(\bm{x}_{\prop})}{D_{0}^\prime \, x_1 \cdots x_n} \,.
	\end{align}
	To show that the above integrand indeed belongs to the sub-sectors, we need to demonstrate that
	\begin{equation}
	A^{(0)}= \frac{D_{0}^{\prime}(x_2,x_7)-D_{1}(\bm{x}_{\prop})}{D_{0}^{\prime}(x_2,x_7)} \,,\quad A^{(i)}=\frac{{N^{(i)}_{0}}^{\prime}(x_2,x_7)-N^{(i)}_{1}(\bm{x}_{\prop})}{D_{0}^{\prime}(x_2,x_7)}
	\end{equation}
	will cancel some denominator factor $x_i$. By construction, we see that $A^{(0)}$ takes the form
	\begin{equation}
		A^{(0)}=\frac{2sx_7-x_7^2}{s^2}+F(x_1,x_3,x_4,x_5,x_6;x_2,x_7) \, ,
	\end{equation}
	where the function $F$ vanishes in the limit $(x_1,x_3,x_4,x_5,x_6) \to 0$. Hence, each term in $F$ must cancel at least one of $x_1,x_3,x_4,x_5,x_6$ in the denominator, while the first term of $A^{(0)}$ cancels $x_7$ in the denominator. Hence, the integrand $I_r^{(0)}$ belongs to the sub-sectors. The integrands $I_r^{(i)}$ are similar. Therefore, we have demonstrated that in this more complicated example, our method still works.

	\subsection{Double-box families with massive propagators}

	\begin{figure}[t]
		\centering
		\includegraphics[width=0.6\textwidth]{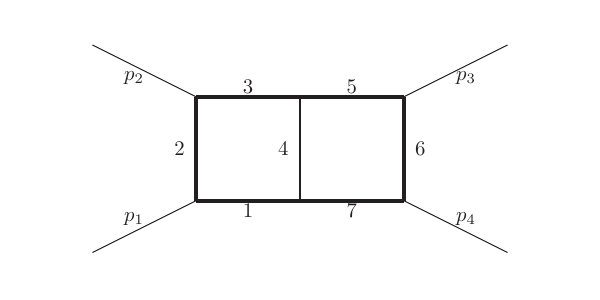}
		\caption{The outer-massive double-box family. Thick lines are massive propagators with the same mass $m$.}\label{fig:outermassivedb}
	\end{figure}
	
	The discussions in the previous subsection can also be extended to double-box families with massive propagators. In Fig.~\ref{fig:outermassivedb}, we depict the outer-massive double-box topology where the thick lines represent massive propagators with the same mass $m$. There are three master integrals in the top sector. For the top-sector ISP reduction, we can borrow a lot from the massless case. We choose the variables $x_i$ to be the same as Eq.~\eqref{eq:propagators_masslessdb}, and the $u$-function is hence unchanged. The propagator denominators are now
	\begin{equation}
	x_1-m^2 \,,\; x_2-m^2 \,,\; x_3-m^2 \,,\; x_4 \,,\; x_5-m^2 \,,\; x_6-m^2 \,,\; x_7-m^2 \,.
	\end{equation}
	With the above denominators, we take the integrand basis $\hat{e}_i$ for the top-sector ISP reduction according to \eqref{eq:doublebox_firstbasis}. From the exchange symmetry, we still have $\bra{e_3} = \bra{e_4} = \bra{e_5}$, but they are now independent of $\bra{e_1}$ in the outer-massive case. Hence, we can choose the top-sector master integrals to be $\bra{e_1},\bra{e_2},\bra{e_3}$.
	
	The top-sector ISP reduction is actually not affected by introducing the internal masses, since the propagators only serve as overall factors of the integrals. Therefore, the reduction coefficients $\tilde{c}_i$ are the same as in Eq.~\eqref{eq:doubleboxfullred}. However, the maximal-cut limit now becomes
	\begin{equation}
	x_4\rightarrow 0 \,,\quad (x_1,x_2,x_3,x_5,x_6,x_7) \rightarrow m^2 \,.
	\end{equation}
	The reduction coefficients $c_1$, $c_2$ and $c_3$ are given by the limits of $\tilde{c}_1$, $\tilde{c}_2$ and $\tilde{c}_3+\tilde{c}_4+\tilde{c}_5$, respectively. The results read
	\begin{equation}
		\begin{aligned}
			c_1&=\frac{-2 m^4 s+2 m^2 s^2+6 m^2 s t+8 m^2 t^2-s^2 t-2 s t^2}{2s(1-2\epsilon)}  \\
			&+\frac{\epsilon  \left(6 m^4 s+8 m^4 t-m^2 s^2+4 m^2 s t+8 m^2 t^2-2 s^2 t-2 st^2\right)}{s(1-2\epsilon)} \, , \\
			c_2&=\frac{-8 m^2 s \epsilon +2 m^2 s-8 m^2 t \epsilon +3 s^2 \epsilon +s t}{s(1-2\epsilon)} \, , \\
			c_3&=\frac{\left(4 m^2-s\right) (s+2 t) \left(4 m^2 s+4 m^2 t-s t\right)}{2s(1-2\epsilon)} \, .
		\end{aligned}
	\end{equation}

	\begin{figure}[t!]
		\centering
		\includegraphics[width=0.6\textwidth]{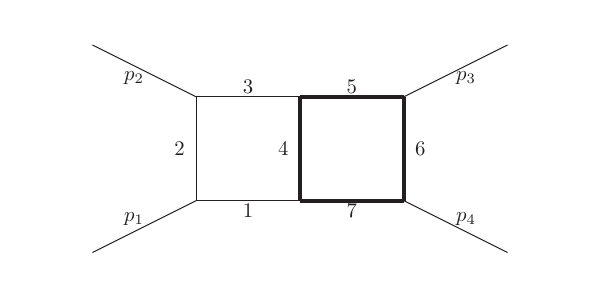}
		\caption{The inner massive double box. Thick lines are massive propagators with the same mass $m$.}\label{fig:innermassivedb}
	\end{figure} 
	
	The situation is slightly more complicated for the inner-massive double-box family, depicted in Fig.~\ref{fig:innermassivedb}. There are four master integrals in the top sector. However, with the integrands in \eqref{eq:doublebox_firstbasis}, $I_3=I_4$ are reducible to $I_1$ under maximal cut:
	\begin{equation}
		\lbra{e_3}=\lbra{e_4}=\frac{1+2\epsilon}{s}\lbra{e_1} + \text{sub-sector integrals} \, .
	\end{equation}
	 Hence, the integrands in \eqref{eq:doublebox_firstbasis} are not enough to serve as a basis for the maximal-cut reduction. This can be easily remedied by introducing a different integrand basis, e.g.:
	\begin{align}\label{eq:diffbasis}
			\tilde{e}_1 &= \tilde{I}(1,1,1,1,1,1,1,0,0) \,,\; \tilde{e}_2 = \tilde{I}(1,1,1,1,1,1,1,-1,0) \,,\; \tilde{e}_3 = \tilde{I}(1,1,1,1,1,1,1,-2,0) \,, \nonumber \\ 
			\tilde{e}_4 &= \tilde{I}(1,1,1,1,1,1,1,0,-1)\,,\; \tilde{e}_5 = \tilde{I}(1,1,1,1,1,1,1,-1,-1) \, .
	\end{align}

	\subsection{The reduction in the sub-sectors of the massless double-box family}
	
	We now consider the reduction in the sub-sectors of the massless double-box family, after subtracting the top-sector components. Unlike the sunrise and banana families whose sub-sectors simply consist of products of one-loop integrals, the sub-sectors of double-box families are non-trivial. Here, new difficulties may arise which require extra treatment before applying the top-down reduction recursively.
	
	Let us demonstrate the situation using a simple example. Take Eq.~\eqref{eq:diffbasis} as the basis for the massless double-box family. Their integrands in the regular form are
	\begin{equation}\label{eq:dbbasis2}
		\begin{aligned}
			\hat{e}_{1}&= \frac{1}{x_1x_2x_3x_4x_5x_6x_7} \,,\; \hat{e}_2=\frac{x_8}{x_1x_2x_3x_4x_5x_6x_7} \,,\; \hat{e}_{3}=\frac{x_8^2}{x_1x_2x_3x_4x_5x_6x_7} \,, \\
			\hat{e}_{4}&= \frac{1}{x_1x_2x_3x_4x_5x_6x_7}\frac{N}{2P_{1}} \,,\; \hat{e}_{5}=\frac{x_8}{x_1x_2x_3x_4x_5x_6x_7}\frac{N}{2P_{1}} \, ,
		\end{aligned}
	\end{equation}
	where the polynomial $P_1$ in the denominator is defined in Eq.~\eqref{eq:dbtopbaikov}. It comes from integrating out $x_9$ from the standard representation. The explicit expression for the numerator $N$ is ($s$ has been set to 1 for simplicity)
	\begin{multline}
			N=t z_2 z_5 z_8-2 t z_2 z_7-z_4 \left(-t z_5+(2 t+1) z_8+z_6\right)+z_3\left(-t z_5^2+z_5 \left((t+1) z_8+z_6\right)\right. \\
			\left.+2 t z_7-z_8\left(z_6+z_8\right)\right)+2 z_1 z_8^2-z_2 z_8^2-2 z_1 z_5 z_8+2 z_2 z_5z_8 +z_2 z_6 z_8+2 z_1 z_7-2 z_2 z_7 \, ,
	\end{multline}
	where the variables $z_i$ are defined in Eq.~\eqref{eq:zvariables}.

    Suppose that we now want to reduce the integral $I(1,1,1,1,1,1,1,-3,0)$, whose integrand is $\hat{\varphi} = x_8^3 / (x_1x_2x_3x_4x_5x_6x_7)$. Performing the top-sector ISP reduction and subtracting the top-sector components, we arrive at the subtracted integrand of the form
	\begin{equation}
		\hat{\varphi}_r = -\frac{x_8[2(x_1-2s)x_8^4]}{s^2P_{1}x_2x_3x_4x_5x_6x_7} + \cdots \, .
	\end{equation}	
	Apparently, the first term in the above belongs to the sub-sector $\{0,1,1,1,1,1,1,0,0\}$, but with an extra polynomial $P_1$ in the denominator. Since $x_1$ is now an ISP in this sub-sector, we may integrate it out and arrive at a lower representation defined by
	\begin{equation}\label{eq:dbsubsectorrep}
		\begin{aligned}
			u_{\setminus 1} &\propto P_{11}^{-\epsilon}P_{12}^{-1/2+\epsilon}P_{13}^{-1/2-\epsilon} \, , \\
			P_{11}&=4G(l_1-p_1,l_2-p_1,p_2) \,, \; P_{12}=-4G(l_2-p_1,p_2) \,, \; P_{13}=16G(l_2,p_1,p_2,p_3) \, ,
		\end{aligned}
	\end{equation}
	where $u_{\setminus 1}$ denotes the $u$-function obtained by integrating out $x_1$. This new representation has one variable less and is easier for calculating intersection numbers. However, the denominator $P_1$ is still present in $\hat{\varphi}_r$ after integrating out $x_1$. The regulator for $P_1$ (which is present in $u$ of \eqref{eq:dbtopbaikov}) is now absent in $u_{\setminus 1}$, since all three factors $P_{11},P_{12},P_{13}$ are different from $P_1$. This means that $P_1$ becomes a relative boundary in this new representation, which destroys one of the benefits of our top-down approach. This is a common problem of our approach in the top-down reduction at two loops and beyond. To proceed, we need to get rid of the denominator $P_1$ before integrating out $x_1$. This can be achieved by IBP relations at the level of the representation \eqref{eq:dbtopbaikov}. It is an interesting question to find an efficient way to perform this kind of transformations.

	\section{Conclusion and Outlook}\label{sec:conclusion}
	
	In this paper, we have further surveyed the recursive structure of Baikov representations \cite{Jiang:2023qnl}, focusing on its application to the reduction of Feynman integrals. We have outlined a systematic approach to perform the top-down reduction using intersection theory of Feynman integrals. Our method completely avoids the introduction of extra regulators for relative boundaries. We also employ the recursive structure to minimize the number of integration variables involved in the intersection theory. These two improvements significantly simplify the computation of intersection numbers. In particular, we find that, for one-loop reductions, there is no need to compute any intersection number at all. We also demonstrate our method using the two-loop sunrise and the three-loop banana families.
	
	A key point in the top-down reduction method is to transform the top-sector subtracted integrands to the sub-sectors. To this end we have introduced the concept of the top-sector ISP reduction. Roughly speaking, we keep the propagator denominators un-integrated, and treat them as constants when performing the IBP reduction. This is used to construct integrands which are IBP equivalent to zero, but when adding them to the top-sector subtracted integrands, the results are manifestly sub-sector integrals. In this work, we perform the top-sector ISP reduction within the framework of intersection theory, but the idea can be easily adopted in other IBP frameworks.
	
	Our method may also be applied as intermediate steps in the construction of canonical bases using the method of intersection theory \cite{Chen:2020uyk, Chen:2022lzr}. In particular, a common complication in both the top-down reduction and the canonical-bases construction is how to transform the integrands with polynomial denominators to ones without, as explained in Sec.~\ref{sec:reductionm}. Only those integrands without extra denominators other than those from propagators explicitly correspond to Feynman integrals. The presence of these extra denominators also spoils the naive application of the recursive reduction approach. It is an interesting question deserving further investigations in the future.
	
	\acknowledgments
	We thank Yanqing Ma and Xin Guan for useful discussions.
	This work was supported in part by the National Natural Science Foundation of China under Grant No. 12375097, 11975030, 12347103, 12047503 and 12225510, and the Fundamental Research Funds for the Central Universities.
	
	\appendix
	
	\section{Reducible sectors and zero sectors}\label{app:dimensionshift}
	
	In Eq.~\eqref{eq:one_loop_dim_shift}, we rewrite a one-loop integral in a reducible sector as sub-sector integrals with a modified $u$-function $P(\bm{x})^\gamma \to P(\bm{x})^{\gamma-1}$, corresponding to a shifted dimension $d \to d-2$ (recall that $\gamma = d/2 + \cdots$). We then need to bring these integrals back to $d$-dimensions. The dimensional recurrence relations are well-understood and are implemented in \texttt{LiteRed} \cite{Lee:2013mka}. To be self-contained, we show here how to construct the necessary relations without resorting to external tools. In the meantime, we also show how to identify zero sectors in which all integrals vanish.
	
	Recall that the one-loop Baikov polynomial $P(\bm{x})$ is a a quadratic polynomial of Baikov variables: $P(\bm{x}) \equiv G(l_1,p_1,\ldots,p_{E})$. For reducible sectors, we have $P(\bm{x}=\bm{0})=0$. This means that the constant term is zero in this polynomial. If the linear terms of $\bm{x}$ are also absent, $P(\bm{x})$ becomes a homogeneous polynomial and this sector can be identified as a zero sector. The reason is that any integral in this sector has an overall scaling behavior under the transformation $\bm{x} \to \lambda \bm{x}$:
	\begin{equation}
		I=\int P(\bm{x})^{\gamma} \prod_{i=1}^{n}\frac{\nd x_{i}}{x_i^{a_i}} \to \lambda^{2\gamma + n-\sum_{i=1}^{n}a_{i}} I \, ,
	\end{equation}
	where $\lambda$ is a non-zero constant. Since $\gamma$ depends linearly on the dimensional regulator $\epsilon$, the power of $\lambda$ is non-zero for any set of integer powers $\{a_i\}$. Hence one can conclude that $I$ is a scaleless integral that vanishes in dimensional regularization.
	
	We can now consider a $P(\bm{x})$ with some linear terms. Without loss of generality, we assume that the linear term of $x_1$ is non-zero, and write $P(\bm{x})$ as
	\begin{equation}
		P(\bm{x})=x_1\left(c+d_{1}x_{1}+\sum_{i=2}^{n}d_{i}x_{i}\right)+\cdots \, ,
	\end{equation} 
	where terms in the ellipsis are independent of $x_1$. We then consider the dimensional recurrence relations. For integrals with a raised power of $P(\bm{x})$ (e.g., in $d+2$ dimensions), we can trivially rewrite them as
	\begin{equation}
		\int P(\bm{x})^{\gamma+1} \prod_{i=1}^{n} \frac{\nd x_{i}}{x_i^{a_i}} = \int P(\bm{x})^{\gamma} \left( P(\bm{x})\prod_{i=1}^{n} \frac{\nd x_{i}}{x_i^{a_i}} \right) .
	\end{equation}
	The right-hand side is an integral in $d$ dimensions with numerators. Integrals with a lowered power of $P(\bm{x})$ (e.g., in $d-2$ dimensions, as we encountered in Eq.~\eqref{eq:one_loop_dim_shift}), are slightly more difficult to deal with. We consider the integral
	\begin{equation}
		I_1 = \int P(\bm{x})^{\gamma-1} \prod_{i=1}^{n} \frac{\nd x_{i}}{x_i^{a_i}} \, .
	\end{equation}
	Using integration-by-parts with respect to $x_1$, we have
	\begin{equation}
		\begin{aligned}
			\int P(\bm{x})^{\gamma} \left( \prod_{i=2}^{n} \frac{\nd x_{i}}{x_i^{a_i}} \right) \nd x_1 \frac{\partial}{\partial x_{1}}\frac{1}{x_1^{a_1}}  + \gamma \int P(\bm{x})^{\gamma-1} \left( \prod_{i=1}^{n} \frac{\nd x_{i}}{x_i^{a_i}} \right) \left( c+2d_{1}x_{1}+\sum_{i=2}^{n}d_{i}x_{i} \right) = 0 \, .
		\end{aligned}
	\end{equation}
	From the above we can solve $I_1$ as
	\begin{equation}
		I_{1} = \frac{a_1}{c\gamma} \int P(\bm{x})^{\gamma} \frac{\nd x_1}{x_1^{a_1+1}} \prod_{i=2}^{n} \frac{\nd x_{i}}{x_i^{a_i}} - \frac{1}{c} \int P(\bm{x})^{\gamma-1}  \left( \prod_{i=1}^{n} \frac{\nd x_{i}}{x_i^{a_i}} \right) \left( 2d_{1}x_{1}+\sum_{i=2}^{n}d_{i}x_{i} \right) .
	\end{equation}	
	The first term is an integral in $d$ dimensions, while the second term consists of integrals in $d-2$ dimensions but with a reduced power of propagators. Applying the above procedure recursively for the second term, one can transform it to a sum of $d$-dimensional integrals and $(d-2)$-dimensional sub-sector integrals. These sub-sector integrals can be dealt with similarly, after integrating out the ISPs. In the end, we can write $I_1$ as a linear combination of $d$-dimensional integrals, as promised.

	\section{Explicit expressions for sunrise and banana families}
	
	In this Appendix, we provide explicit expressions for certain integrands and reduction coefficients appeared in Sec.~\ref{sec:reduction}.
	
	\subsection{The unequal-mass sunrise family}\label{app:sunrise}
	
	Here we list the explicit expressions for $N_{1}^{(i)}$ in Eq.~\eqref{eq:fullreduction2}:
	\begin{align}
		N_{1}^{(1)}&=(2\epsilon -1)\left(s-s_1\right) s_1 \left(s^2-2 \left(s_1+s_2+s_3\right)s+s_1^2+s_2^2+s_3^2+6 s_2 s_3-2 s_1 \left(s_2+s_3\right)\right) \, , \nonumber \\
		N_{1}^{(2)}&=(2\epsilon -1)\left(s-s_2\right) s_2 \left(s^2-2 \left(s_1+s_2+s_3\right)s+s_1^2+\left(s_2-s_3\right){}^2-2 s_1 \left(s_2-3 s_3\right)\right)\, , \nonumber \\
		N_{1}^{(3)}&=-\frac{s^4 \epsilon }{2}+s^3 \left(2 s_1 \epsilon +2 s_2 \epsilon +4 s_3\epsilon -s_3\right) \, \nonumber \\
		&+s^2 \left(-3 s_1^2 \epsilon -2 s_2 s_1 \epsilon -6s_3 s_1 \epsilon -3 s_2^2 \epsilon -9 s_3^2 \epsilon -6 s_2 s_3 \epsilon+2 s_3 s_1+3 s_3^2+2 s_2 s_3\right) \nonumber \\
		&+s \left(2 s_1^3 \epsilon -2 s_2 s_1^2\epsilon -2 s_2^2 s_1 \epsilon -2 s_3^2 s_1 \epsilon +32 s_2 s_3 s_1\epsilon +2 s_2^3 \epsilon +8 s_3^3 \epsilon -2 s_2 s_3^2 \epsilon -s_3s_1^2\right. \nonumber \\ 
		&\left.-6 s_2 s_3 s_1-3 s_3^3-s_2^2 s_3\right)+\frac{1}{2} \left(-s_1^4\epsilon+4 s_2 s_1^3 \epsilon +4 s_3 s_1^3 \epsilon -6 s_2^2 s_1^2\epsilon -10 s_3^2 s_1^2 \epsilon \right. \nonumber \\
		&-4 s_2 s_3 s_1^2 \epsilon +4 s_2^3 s_1\epsilon +12 s_3^3 s_1 \epsilon -28 s_2 s_3^2 s_1 \epsilon -4 s_2^2 s_3s_1 \epsilon -s_2^4 \epsilon -5 s_3^4 \epsilon +12 s_2 s_3^3 \epsilon \nonumber \\
		&\left. -10s_2^2 s_3^2 \epsilon +4 s_2^3 s_3 \epsilon +2 s_3^2 s_1^2-4 s_3^3 s_1+12s_2 s_3^2 s_1+2 s_3^4-4 s_2 s_3^3+2 s_2^2 s_3^2\right) \, , \nonumber \\
		N_{1}^{(4)}&=\frac{1}{2}(2 \epsilon -1) (3 \epsilon -2) \left(s^3-3 \left(s_1+s_2+s_3\right) s^2 \right. \nonumber \\
		&+\left(3 s_1^2+2\left(s_2+s_3\right) s_1+3 s_2^2+3 s_3^2+2 s_2 s_3\right)s  \nonumber \\ 
		&\left.-s_1^3+s_1^2\left(s_2+s_3\right)-\left(s_2-s_3\right){}^2 \left(s_2+s_3\right)+s_1\left(s_2^2-10 s_3 s_2+s_3^2\right)\right) \, .
	\end{align}
	
	\subsection{The 3-loop banana family}\label{app:banana}
	The coefficients $\tilde{c}_{i}$ in \eqref{eq:bananatop} are:
	\begin{align}
		\tilde{c}_1=&\frac{1}{12(-1+\epsilon)^2}\left[30 s \epsilon ^2-18 s_1 \epsilon ^2-6 s_2 \epsilon ^2-6 s_3 \epsilon ^2-6 s_4\epsilon ^2-53 s \epsilon +23 s_1 \epsilon +7 s_2 \epsilon +7 s_3 \epsilon \right. \nonumber \\
		&   \hspace{70pt}\left.+7 s_4 \epsilon +24 s-6 s_1-2 s_2-2 s_3-2 s_4\right] \, , \nonumber \\
		\tilde{c}_2=&\frac{(-5+6\epsilon)(s-s_1)s_1}{6(-1+\epsilon)^2} \, , \nonumber \\
		\tilde{c}_3=&\frac{s_2 \left(2 s \epsilon -4 s_1 \epsilon -2 s_2 \epsilon -4 s_3 \epsilon -4 s_4\epsilon -s+4 s_1+s_2+2 s_3+2 s_4\right)}{6(-1+\epsilon)^2}\, , \nonumber \\
		\tilde{c}_4=&\frac{s_3 \left(2 s \epsilon -4 s_1 \epsilon -4 s_2 \epsilon -2 s_3 \epsilon -4 s_4\epsilon -s+4 s_1+2 s_2+s_3+2 s_4\right)}{6(-1+\epsilon)^2}\, , \nonumber
    \end{align}
    \begin{align}
		\tilde{c}_5=&\frac{-1}{12(-1+\epsilon)^2}\left[s^2 \epsilon -2 s_1 s \epsilon -2 s_2 s \epsilon -2 s_3 s \epsilon -6 s_4 s\epsilon +s_1^2 \epsilon +s_2^2 \epsilon +s_3^2 \epsilon +5 s_4^2 \epsilon -2 s_1 s_2 \epsilon \right. \nonumber \\
		&\left.  -2 s_1 s_3 \epsilon +6 s_2 s_3 \epsilon +6 s_1 s_4\epsilon +14 s_2 s_4 \epsilon +14 s_3 s_4 \epsilon +2 s_4 s\!-\!2 s_4^2\!-\!8 s_1s_4\!-\!4 s_2 s_4\!-\!4 s_3 s_4\right] \, ,\nonumber \\
		\tilde{c}_6=&0, \, \tilde{c}_7 =0, \tilde{c}_8=\frac{s_2 s_4\left(s+s_1-s_2-3 s_3-s_4\right)}{3(-1+\epsilon)^2}, \, \nonumber \\
		\tilde{c}_9=&\frac{s_2 s_3 \left(s+s_1-s_2-s_3-3 s_4\right)}{3(-1+\epsilon)^2}\, , \tilde{c}_{10}=\frac{s_3 s_4 \left(s+s_1-3 s_2-s_3-s_4\right)}{3(-1+\epsilon)^2} \, , \nonumber \\
		\tilde{c}_{11}=&\frac{-s_4}{6(-1+\epsilon)^2}\left[s^2-2 s_1 s-2 s_2 s-2 s_3 s-2 s_4 s+s_1^2+s_2^2+s_3^2+s_4^2-2 s_1 s_2-2 s_1s_3 \right. \nonumber \\
		& \hspace{65pt}\left. +6 s_2 s_3-2 s_1 s_4+6 s_2 s_4+6 s_3 s_4\right] \, .
	\end{align}

\bibliographystyle{apsrev4-1}
\bibliography{ref_inspire,ref}

\end{document}